\documentclass[reprint,amsmath,amssymb,aps,pra,floatfix,superscriptaddress]{revtex4-2}

\usepackage{graphicx}
\usepackage{float}
\usepackage{soul}
\usepackage[normalem]{ulem}

\usepackage{amssymb}
\usepackage{amsmath}
\usepackage[unicode=true, breaklinks=false, pdfborder={0 0 1}, backref=false, colorlinks=true, linkcolor=blue, citecolor=blue, urlcolor=blue]{hyperref}
\usepackage{siunitx}
\usepackage{color}
\usepackage[usenames,dvipsnames]{xcolor}
\usepackage[mathscr]{eucal}
\usepackage{mathtools}
\usepackage[percent]{overpic}

\newcommand{\new}[1]{ {\color{black} #1} }

\def\hatvPi{{\boldsymbol{\hat\varPi}}}
\def\tildevPi{{\boldsymbol{\widetilde\varPi}}}
\def\vPi{\boldsymbol{\varPi}}
\def\hatvP{{\boldsymbol{\hat P}}}
\def\hatvA{\boldsymbol{\hat A}}

\def\hatvp{\boldsymbol{\hat p}}
\def\hatvr{\boldsymbol{\hat r}}

\def\keps{{\boldsymbol{k}\eps}}
\def\pmu{{\boldsymbol{p}\mu}}
\def\qmu{{\boldsymbol{q}\nu}}

\def\dens{\xi}

\newcommand{\da}{\dagger}
\newcommand{\la}{\langle}
\newcommand{\ra}{\rangle}
\newcommand{\id}{\ensuremath{\mathbb I}}

\newcommand{\pmat}[1]{\begin{pmatrix} #1 \end{pmatrix}}

\renewcommand\vec[1]{\boldsymbol{#1}}
\renewcommand\vr{\vec{r}}
\newcommand\vk{\vec{k}}
\newcommand\va{\vec{a}}
\newcommand\vp{\vec{p}}

\newcommand\vd{\vec{d}}
\newcommand\vD{\vec{D}}
\newcommand\vA{\vec{A}}

\newcommand\vE{\vec{E}}
\newcommand\vB{\vec{B}}
\newcommand\vS{\vec{S}}
\newcommand\vF{\vec{F}}
\newcommand\vn{\vec{n}}
\newcommand\vx{\vec{x}}
\newcommand\vg{\vec{g}}

\newcommand{\vrho}{\vec{\rho}}
\newcommand{\vtheta}{\vec{\theta}}
\newcommand{\vvtheta}{\vec{\Theta}}
\newcommand{\valpha}{\vec{\alpha}}
\newcommand{\pulsewidth}{\tau}
\newcommand{\lambdain}{\lambda^{\rm in}}
\newcommand{\kin}{k^{\rm in}}
\newcommand{\vkin}{\boldsymbol{k}^{\rm in}}
\newcommand{\omegain}{\omega^{\rm in}}

\newcommand{\Op}[1]{\hat{#1}}

\newcommand{\oa}{\Op{a}}
\newcommand{\ova}{\Op{\va}}
\newcommand{\ob}{\Op{b}}
\newcommand{\oH}{\Op{H}}
\newcommand{\oM}{\Op{M}}
\newcommand{\oL}{\Op{L}}

\newcommand{\oU}{\Op{U}}
\newcommand{\oD}{\Op{D}}

\newcommand{\ovp}{\Op{\vec{p}}}
\newcommand{\ovr}{\Op{\vec{r}}}

\newcommand{\eps}{\varepsilon}
\newcommand{\vac}{\mathrm{vac}}
\newcommand{\tr}{\mathrm{tr}}
\newcommand\ve{\mathbf{e}} 

\newcommand{\eqreff}[1]{(\ref{#1})}
\newcommand{\reff}[1]{\ref{#1}}

\newcommand{\added}[1]{{\color{black} #1}}
\newcommand{\addded}[1]{{\color{black} #1}}

\begin{document}

\title{Quantum Limits of Position and Polarizability Estimation in the Optical Near Field}

\author{Lukas Kienesberger}
  \affiliation{University of Vienna, Faculty of Physics, VCQ,
A-1090 Vienna, Austria}
  \affiliation{University of Vienna, Max Perutz Laboratories,
Department of Structural and Computational Biology, A-1030 Vienna,
Austria}
  \affiliation{Department of Physics, Ludwig-Maximilians-Universit\"at M\"unchen, Theresienstraße 37, D-80333 M\"unchen, Germany}

\author{Thomas Juffmann}
  \affiliation{University of Vienna, Faculty of Physics, VCQ,
A-1090 Vienna, Austria}
  \affiliation{University of Vienna, Max Perutz Laboratories,
Department of Structural and Computational Biology, A-1030 Vienna,
Austria}

\author{Stefan Nimmrichter}
\affiliation{Naturwissenschaftlich-Technische Fakult\"at, Universit\"at Siegen, Walter-Flex-Stra\ss e 3, 57068 Siegen, Germany}
\date{\today}

\begin{abstract}
Optical near fields are at the heart of various applications in sensing and imaging. We investigate dipole scattering 
as a parameter estimation problem
and show that optical near-fields carry more information about the location and the polarizability of the scatterer than the respective far fields. This increase in information originates from, and occurs simultaneously with, the scattering process itself. Our calculations also yield the far-field localization limit for dipoles in free space. 
\end{abstract}

\maketitle

Near fields have applications ranging from
nanofabrication~\cite{Paik2020Near-fieldPhotomask} to sensing~\cite{Gordon2019INVITEDTweezers, Mandal2022ProgressReview} and imaging~\cite{Pohl1984Optical/20, NovotnyB.2012PrinciplesNano-optics}. They enable enhanced, highly localized interactions and label-free imaging at a spatial resolution beyond the diffraction limit, with illumination wavelengths from the optical to the radio-frequency range. 

With recent advances in far-field label-free super-resolution imaging~\cite{Pushkina2021SuperresolutionField}, the question arises whether there is a fundamental advantage of operating in the near-field regime. We approach this question as a parameter estimation task. In optical imaging, information about a parameter of interest is encoded into the state of the probing light. It is quantified by the quantum Fisher information (QFI). Information retrieved in a specific measurement on that probe state is quantified by the Fisher information (FI)~\cite{Bouchet2021, Ly2017AInformation, Liu2020QuantumEstimation, Toth2014QuantumPerspective, Safranek2017}. These two quantities determine the (Quantum) Cram\'{e}r-Rao bounds (QCRB) on the minimum parameter estimation variance achievable for a specific probe state or measurement, respectively.

\added{Based on this framework,} 
one can analyze and improve measurement techniques in practice. The localization precision was optimized in fluorescence microscopy~\cite{Chao2016, Shechtman2014OptimalImaging, Backlund2018FundamentalStatistics, Balzarotti2017NanometerFluxes} and interferometric scattering microscopy~\cite{Dong2021FundamentalPhotometry}, the phase estimation precision in phase microscopy and holography~\cite{Bouchet2021FundamentalMicroscopes, Koppell2021InformationImaging}, and the lifetime estimation precision in fluorescence lifetime microscopy~\cite{Bouchet2019Cramer-RaoMicroscopy,Mitchell2022QuantumLifetimes}. One could also optimize measurements in challenging scenarios, e.g., when an object of interest is embedded in a highly scattering medium~\cite{Bouchet2021}.
These ideas were recently extended to \added{quantum optical scattering}~\cite{hüpfl2023continuity, Albarelli2023FundamentalEstimation} and electron microscopy, where dose-induced damage limits the number of \added{electron-probe interactions~\cite{Koppell2022TransmissionLimit, Dwyer2023QuantumMicroscopy}.}

Here, we consider optical microscopy and calculate the (Q)FI and (Q)CRB regarding the position and polarizability of a scatterer both in the near and far field. The (Q)CRB on localization are relevant for tracking~\cite{Taylor2019InterferometricMembrane} and imaging, while those on polarizability are relevant for sizing and mass photometry applications~\cite{Young2018QuantitativeMacromolecules}.
We first describe the scattering process classically and \addded{calculate the CRB based on a standard Poissonian detection model. At distances closer than the probe wavelength, we find a CRB that is significantly lower (i.e.~better) than in the far field where the ideally achievable uncertainties for position and polarizability estimation are constant. In the near field, they can improve, respectively, with the third and the second power of the detector distance.}

\addded{Our phenomenological assessment of the CRB, however, extrapolates an ideally resolving and non-invasive photodetector model from the far to the near field. For a more fundamental bound, 
we move on to solve the full time-dependent quantum scattering problem for a free scatterer. We find that the QFI contained in the quantum state of the field} 
is significantly enhanced while the probe-sample interaction takes place. \addded{In the far field, the resulting QCRB} bounds neither the CRB nor the QCRB in the near field, and near-field measurements can, therefore, be more precise than \emph{any} (coherent) far-field measurement performed with the same probe light. 

\begin{figure}
    \centering
    \includegraphics[width=\linewidth]{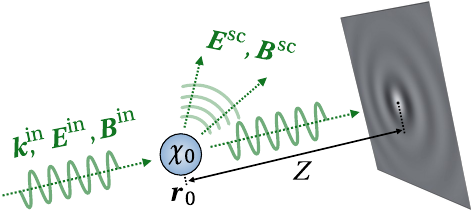}
    \caption{Sketch of the planar detector setup for light scattered off a dipole polarizability $\chi_0$ at position $\vr_0$. The detector plane is orthogonal to the wave vector of the incident light \added{($\vk^{\rm in}$, $\boldsymbol E^{\rm in}$, $\boldsymbol B^{\rm in}$)}, at a distance $Z$ from the dipole. The scattered light is denoted by ($\boldsymbol E^{\rm sc}$, $\boldsymbol B^{\rm sc}$).
    }
    \label{fig:sketch}
\end{figure}

\textit{Dipole scattering model.---}We consider the setting sketched in Fig.~\ref{fig:sketch}: a dipole scatterer located at $\vr_0$ is illuminated by coherent, linearly polarized, and monochromatic light propagating along the $z$-axis with wave vector $\vk^{\rm in} = \kin \ve_z $. Its amplitude is given by $E^{\rm in}$ and its polarization by $\ve_x$.
The scatterer's linear response to this field is characterized by a scalar dipole polarizability $\chi_0$. The task is to estimate $\chi_0$ and the position $\vr_0 = (x_0,y_0,z_0)$ by measuring the light in a position-resolving photo-detector placed in the near or far field of the scatterer.

We will approach the estimation task in two ways corresponding to different degrees of scrutiny. The first approach is phenomenological: we treat the incident light as a plane-wave field of wavelength $\lambdain = 2\pi/\kin$, $\vE^{\rm in} (\vr,t) = E^{\rm in} \ve_x e^{i\kin(z-ct)}$ and $\vB^{\rm in} = \ve_z \times \vE^{\rm in}/c$, and the scatterer as the classical induced Hertz dipole $\vd(t) = 2\epsilon_0 \chi_0 \vE^{\rm in} (\vr_0,t)$ that oscillates at the light frequency $\omegain = c\kin$. Information about the scatterer's position $\vr_0$ and polarizability $\chi_0$ is broadcast to an ideal position-resolving (and backaction-free) photodetector through dipole radiation,
\begin{align}\label{scattered_field}
  \boldsymbol E^{\rm sc}(\boldsymbol r, t)
  &= \frac{(\kin)^3 \chi_0 E^{\rm in}}{2\pi} e^{i \kin(\rho+z_0-ct)}
  \left[
  \frac{(\ve_{\vrho} \times \ve_x) \times \ve_{\vrho}}{\kin\rho} \right. \nonumber \\
  &\quad \left. + \frac{\ve_x - 3\ve_{\vrho}(\ve_x \cdot\ve_{\vrho})}{(\kin\rho)^3}
  (i\kin\rho-1)
  \right],
  \\
  \boldsymbol B^{\rm sc}(\boldsymbol r, t)
  &= \frac{i(\kin)^3 \chi_0 E^{\rm in}}{2\pi c} e^{i \kin(\rho+z_0-ct)}
  \frac{\ve_{\vrho} \times \ve_x}{(\kin\rho)^2} (1 - i\kin\rho), \nonumber 
\end{align}
where we define $\vrho = \boldsymbol{r-r}_0$ and $\ve_{\vrho} = \vrho/\rho$. We consider a planar detector surface in the $z=Z$ plane here; see Sec.~II in \cite{Supp} for a hemispherical detector of radius $R$.

Our second approach is a dynamical scattering model: the dipole is a quantum harmonic oscillator of frequency $\omega_0$ aligned with the electric field of the incident light, which is a \added{Gaussian pulse with slowly varying amplitude, $E^{\rm in} (t) = E^{\rm in} e^{-\pi t^2/2\tau^2}$, occupying a narrow frequency band $\Delta \omega \sim 1/\tau$ around $\omegain < \omega_0$.}
In the multipolar  gauge~\cite{Stokes2022ImplicationsElectrodynamics}, the light-matter coupling reduces to the well-known dipole Hamiltonian (see Sec.~III in \cite{Supp}),
\begin{equation}\label{interactionHamiltonian}
    \oH_I = (\ob + \ob^\da) \sum_{\vk,\eps} \sqrt{\frac{\hbar c k}{2\epsilon_0 L^3}} 
    d_0 \dens_{\vk} (\ve_x\cdot\ve_{\keps})
    \left[ \frac{e^{i\vk\cdot\vr_0}}{i} \oa_{\vk \eps} + h.c.\right],
\end{equation}
with $\ob$ the dipole's ladder operator, $d_0 \in \mathbb{R}$ its strength parameter, and $\oa_{\vk \eps}$ the bosonic operators associated to plane-wave modes of wave vectors $\vk$ and transverse polarizations $\ve_{\keps} \perp \vk$ ($\eps=1,2$) in the mode volume $L^3$. We alleviate high-frequency divergences arising from an ideal point dipole by introducing 
\added{the regularization $\dens_{\vk} = 16[4 + (a_0 k)^2]^{-2}$~\cite{Stokes2022ImplicationsElectrodynamics}. As we show in Sec.~VII of \cite{Supp}, this amounts to relaxing the point-dipole approximation to the finite, exponentially localized polarizability density $\dens(\vr) = e^{-2r/a_0}/(\pi a_0^3)$. 
Corrections to the dipole Hamiltonian are negligible as long as the size parameter $a_0$} is much smaller than the populated wavelengths.
Note that we do not truncate the dipole to a two-level system, as this would complicate the calculation and is known to cause problems with gauge invariance~\cite{Stokes2022ImplicationsElectrodynamics}.

\textit{Classical near-field CRB.---}
We start with the phenomenological model and evaluate how well \added{an ideal shot noise-limited detector recording a spatial distribution of photon counts} can resolve the parameters $\vtheta = (\chi_0,x_0,y_0,z_0)$ of the scatterer \added{emitting according to \eqref{scattered_field}}. 

A single detector pixel of area $d A$ at position $\vr$ sees a light intensity $ I (\vr,\vtheta) = \ve_{\vn} \cdot \vS (\vr,\vtheta)$, with $\ve_{\vn}$ the unit vector orthogonal to the pixel surface and $\vS (\vr,\vtheta)$
the time-averaged Poynting vector of the total field at $\vr$; it depends on $\vtheta$ through the dipole field \eqref{scattered_field}. 
\added{Integrated over a measurement time window $\tau$ (e.g., the duration of a narrow-band pulse), an average of $\bar{n} (\vr,\vtheta) = I (\vr,\vtheta) \tau d A /\hbar \omega$ photons are detected. The likelihood to count $n$ photons in each pixel $\vr$ is modeled by a Poisson distribution~\cite{Chao2016}, $p(n|\vr,\vtheta) =  e^{-\bar{n} (\vr,\vtheta)} [\bar{n} (\vr,\vtheta)]^n /n!$.
Assuming independent pixels with no cross-talk, the likelihood for a recorded distribution of photon counts is given by the product of the individual pixels' likelihoods. The overall sensitivity to variations in $\vtheta$ is measured by the FI matrix~\cite{Dong2021FundamentalPhotometry}
\begin{equation}
  \label{eq:FIMPoissonContinuum}
  \mathcal I_{j \ell}(\vtheta) = \frac{\tau}{\hbar\omega} \int_{\text{pixels}} \frac{1}{I(\vr,\vtheta)} \frac{\partial I(\vr,\vtheta)}{\partial \theta_j} \frac{\partial I(\vr,\vtheta)}{\partial \theta_\ell} dA.
\end{equation}
It determines how precisely one can infer the parameter values from a sample of measurement data: the mean-square error of any unbiased estimate of each $\theta_\ell$, $ \ell = 0,1,2,3$, is lower-bounded by $(\Delta \theta_\ell)^2 \geq [\mathcal I^{-1} (\vtheta) ]_{\ell \ell}$, with $\mathcal I^{-1}$ the matrix inverse. This is known as the CRB \cite{van2007parameter}.} 

\begin{figure}
\begin{overpic}[width=\columnwidth]{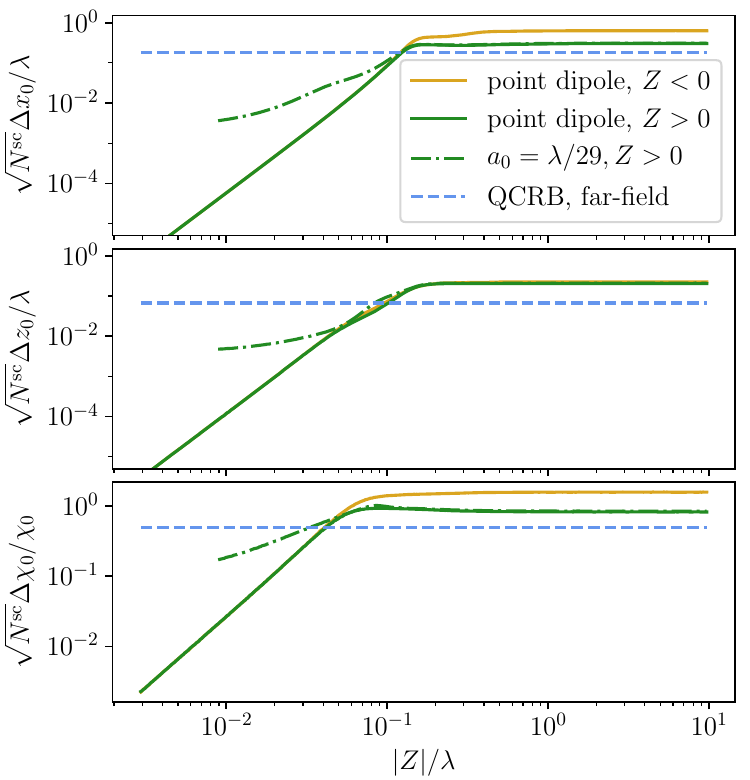}
  \put (18, 86) {\large (a)}
  \put (18, 54) {\large (b)}
  \put (18, 22) {\large (c)}
\end{overpic}
\caption{
Cram\'{e}r-Rao bounds for estimating (a) the $x_0$-position, (b) the $z_0$-position, and (c) the polarizability $\chi_0$ of a dipole scatterer with a  planar detector at varying distance $Z$. \added{The square-shaped detector always covers a solid angle of $1.97\pi$.} We compare forward and backward scattering for a point dipole, a finite-size scatterer, and the far-field quantum CRBs, for a total number of scattered photons $N^{\rm sc}$.}
\label{fig:CRB_plots_planar}
\end{figure}

Figure \ref{fig:CRB_plots_planar} shows the CRB for \added{a large planar detector ($\ve_{\vn}=\ve_z$) covering a fixed solid angle $\Omega = 1.97\pi$} at varying distance $Z$, in an exemplary setting with polarizability $\chi_0 = \SI{13.0}{nm^3}$ at $\lambdain = \SI{1.03}{\mu m}$ and a number $N^{\rm sc} = \sigma_{\rm tot} c\epsilon_0 |E^{\rm in}|^2 \tau/2\hbar\omegain$ of scattered photons, with $\sigma_{\rm tot} = 2(\kin)^4|\chi_0|^2/3\pi$ the total scattering cross section. We plot the CRB for (a) $x_0$, (b) $z_0$, and (c) $\chi_0$ estimation, comparing forward ($Z>0$) and backward scattering ($Z<0$) at a point dipole, as well as forward scattering at a polarization density of size $a_0 = \SI{35}{nm}$; see Sec.~II in \cite{Supp}.
The CRB always saturates to a distance-independent value in the far field, $|Z|\gtrsim \lambdain$. 
Conversely, the dipole fields \eqref{scattered_field} diverge at the scatterer position, and so does the FI, which implies that the CRB would vanish for an ideal detector placed arbitrarily closely. In the intermediate near field not too close to the dipole, $ (4\pi \chi_0)^{1/3} \ll |Z| \ll \lambdain$, we can neglect the contribution $\vE^{\rm sc} \times \vB^{\rm sc*}$ to the Poynting vector, which results in the scaling $\Delta \theta_0 \sim |Z|^2 $ and $\Delta \theta_{\ell\neq 0} \sim |Z|^3$ for the CRB of polarizability and position, respectively. This scaling is seen in the diagrams for $|Z|\lesssim 0.1\,\lambdain$, though finite-size corrections limit the precision when $|Z|\sim a_0$.

The dashed line marks the fundamental QCRB \addded{for far-field detection, based on Eq.~\eqref{FarFieldQFI} below.}
The fact that the saturated bounds on the right of Fig.~\ref{fig:CRB_plots_planar} are worse than the QCRB shows that the specified detection scheme is not optimal for estimating the parameters.
This is in contrast to 
interferometric scattering microscopy~\cite{Taylor2019Review}, coherent bright field microscopy~\cite{Hsieh2018Label-freeImaging}, or dark-field microscopy~\cite{Weigel2014DarkSensitivity}, 
which can reach the QCRB under certain conditions~\cite{Dong2021FundamentalPhotometry}.

\textit{Near-field QCRB.---}
The previous phenomenological model has \addded{three crucial limitations}. Firstly, it assumes a phase-insensitive photo-detector model of a specific geometry. \addded{Any phase information or light that does not reach the detector is not accounted for.} 
Secondly, it over-simplifies the scatterer's response to the probe light by a quasi-instantaneous dipole field throughout the pulse duration. \addded{Thirdly, photon detection is taken to be a separate event from photon scattering, neglecting the possible influence of the detector on the near-field mode structure}. 

\addded{To circumvent these issues, we derive fundamental quantum precision bounds based on the overall information the scatterer broadcasts into the state of the electromagnetic field while it emits radiation. That is,}
\added{we calculate the quantum field state $\varrho (t,\vtheta)$ at every time $t$ during the scattering process and quantify its sensitivity to $\vtheta$-variations in terms of the measurement model-independent QFI matrix $\mathcal J_{j\ell}(\vtheta,t)$ \cite{Liu2020QuantumEstimation, Toth2014QuantumPerspective, Safranek2017}. It was proven that, whatever measurement one performs on $\varrho (t,\vtheta)$ to infer $\vtheta$, the mean-square errors of unbiased parameter estimates obey the QCRB inequalities $(\Delta \theta_{\ell})^2 \geq [\mathcal{J}^{-1}(\vtheta,t)]_{\ell \ell}$
\cite{Liu2020QuantumEstimation}. The QFI, a function of $\varrho (t,\vtheta)$, thus serves as a fundamental precision benchmark that may not always be attainable in a practical measurement.

To obtain $\varrho (t,\vtheta)$ in our case, we must evolve the pure quantum state of field and scatterer unitarily according to the Hamiltonian $\oH = \sum_{\vk,\eps} \hbar c k \, \oa_{\vk \eps}^\da \oa_{\vk \eps} + \hbar\omega_0 \ob^\da \ob + \oH_I$ and then take the partial trace over the scatterer degree of freedom. 
Fortunately, the asymptotic initial state at $t\to-\infty$ is Gaussian: it describes the incident light pulse by a coherent displacement of the mode vacuum with amplitudes $\valpha^{\rm in} = (\alpha^{\rm in}_{\vk\eps})_{\vk,\eps}$ and the scatterer in its ground state.}
Given the linear interaction Hamiltonian \eqref{interactionHamiltonian}, the state remains Gaussian at all times and is therefore fully characterized by the time evolution of its first and second moments in the mode operators, which depend on the parameters $\vtheta$.

At each point in time $t$, the reduced state of the radiation field is determined by a vector of mean coherent amplitudes, $\valpha = \la \ova \ra$ with elements $\alpha_{\vk\eps} (t) = \la \oa_{\vk\eps} (t) \ra$, and by covariance matrix blocks $\Xi  = 2[\la \ova \circ \ova^\da \ra - \valpha \circ \valpha^* ] - \id$ and $\Upsilon = 2[\la \ova \circ \ova \ra - \valpha \circ \valpha ]$ with '$\circ$' denoting the dyadic product. 
\added{The QFI matrix for Gaussian states} 
reads as~\cite{Safranek2017}
\begin{align}\label{GaussianQFI}
    \mathcal J_{j\ell} (\vtheta,t) &= 2 \pmat{\frac{\partial \valpha^*}{\partial \theta_{j}} & \frac{\partial \valpha}{\partial \theta_{j}} } \pmat{ \Xi & \Upsilon \\ \Upsilon^* & \Xi^* }^{-1} \pmat{ \frac{\partial \valpha}{\partial \theta_{\ell}} \\ \frac{\partial \valpha^*}{\partial \theta_{\ell}} }
    + \mathcal V_{j\ell} (\vtheta,t)\nonumber \\
    &\approx 2 \left[ \tfrac{\partial \valpha^*}{\partial\theta_j} \cdot \tfrac{\partial \valpha}{\partial\theta_{\ell}} + c.c. \right] + \mathcal V_{j\ell} (\vtheta,t) \\
    &\quad - 2\pmat{ \frac{\partial \valpha^*}{\partial \theta_{j}} & \frac{\partial \valpha}{\partial \theta_{j}} } \pmat{ \Xi - \id & \Upsilon \\ \Upsilon^* & \Xi^* - \id } \pmat{ \frac{\partial \valpha}{\partial \theta_{\ell}} \\ \frac{\partial \valpha^*}{\partial \theta_{\ell}} }.  \nonumber
\end{align}
In the second line, we expand the inverse of the covariance matrix to first order around the identity matrix, a good approximation for realistic weak scatterers. The lengthy additional term $\mathcal V$ does not depend on the amplitudes $\valpha$ and is thus present even when there is no incident light. It stems from the higher-order effect that the presence of the scatterer squeezes the surrounding mode vacuum, which for realistic light intensities would add only little to the information contained in the $\valpha$-terms in \eqref{GaussianQFI}. Assuming that the parameter estimation is based on coherent amplitude measurements, we can safely ignore $\mathcal V$ in the following.

The Heisenberg time evolution of the field operators under $\oH$ can be solved in a lengthy calculation assuming weak coupling (Sec.~IV in \cite{Supp}). In particular, the mean amplitudes $\valpha(t)$ are linearly related to the incident $\valpha^{\rm in}$, 
\begin{equation}\label{AmplitudeTransformation}
  \alpha_\pmu(t) = \sum_\keps \left( u_{\pmu,\keps} \alpha_\keps^{\rm in} e^{-ick t} + v_{\pmu,\keps} \alpha_\keps^{\rm in *} e^{ick t} \right).
\end{equation}
The transformation coefficients are
\begin{align}\label{eq:uvpk_MainText}
    u_{\pmu,\keps} &= 
    \frac{\sqrt{kp}}{L^3}
    (\ve_\pmu \cdot \ve_x) (\ve_\keps \cdot \ve_x) \dens_{\vp} \dens_{\vk}
    \frac{\chi (ck)e^{-i(\vp-\vk)\cdot\vr_0}}{p-k-i0^+} \nonumber
    \\ &\quad + \delta_{\pmu,\keps},
    \\ \nonumber
    v_{\pmu,\keps} &= -\frac{\sqrt{kp}}{L^3}
    (\ve_\pmu \cdot \ve_x) (\ve_\keps \cdot \ve_x) \dens_{\vp} \dens_{\vk}
    \frac{\chi (ck) e^{-i(\vp+\vk)\cdot\vr_0}}{p+k}. \nonumber
\end{align}
The matrix elements of $\Xi - \id$ and $\Upsilon$ can also be given. However, since they themselves are weak-coupling corrections, their contribution in the last line of \eqref{GaussianQFI} can be safely neglected, as we demonstrate in Sec.~V of \cite{Supp}.

The expression \eqref{AmplitudeTransformation} simplifies greatly in the far field. Introducing the asymptotic output amplitudes $\valpha^{\rm out}$ as $\alpha_{\keps}^{\rm out} = \lim_{t\to\infty} \alpha_{\keps} (t) e^{ickt}$ and treating the $\vk$-modes as a continuum (see Sec.~VI in \cite{Supp}), we arrive at
\begin{align}\label{FarFieldAmplitude}
    \alpha_{\pmu}^{\rm out} = \alpha_{\pmu}^{\rm in} + \frac{i p}{4\pi^2} & \int d^3 k \, \delta (p-k) \chi (ck) e^{i(\vk-\vp)\cdot\vr_0} \\
    &\times \sum_{\eps} (\ve_x\cdot\ve_{\keps}) (\ve_x\cdot\ve_{\pmu}) \alpha^{\rm in}_{\keps}. \nonumber
\end{align}
This amounts to elastic light scattering via a dipole polarizability, described by the linear response function
$\chi(\omega) = d_0^2 \omega_0/\hbar\epsilon_0 (\omega_0^2-\omega^2)$.
Far off resonance, it is approximately constant, $\chi (\omega \ll \omega_0) \approx d_0^2/\hbar\epsilon_0\omega_0 \approx \chi_0$, reconciling the quantum oscillator model with the previous phenomenological description based on the polarizability $\chi_0 = \chi(\omegain)$. Indeed, we show in Sec.~VII of \cite{Supp} that the light field expectation values for monochromatic input match the dipole radiation terms \eqref{scattered_field}.

To leading weak-coupling order in the far field, the QFI matrix \eqref{GaussianQFI} reduces to a diagonal matrix with elements
\begin{equation}\label{FarFieldQFI}
[\mathcal J_{\ell \ell} (\vtheta,\infty)]_{\ell=0}^3 = \frac{8 (\kin)^6 |\chi_0|^2 \Phi }{15\pi} \left[ \frac{5}{(\kin |\chi_0|)^2}, 1, 2, 7 \right] ,
\end{equation}
with $\Phi = (1/L^2) \sum_\keps |\alpha_\keps^{\rm in}|^2 = c\epsilon_0 |E^{\rm in}|^2 \tau/2\hbar\omegain$ the number of incident photons per area. 
Consequently, the far-field precision limits for the polarizability and position of the scatterer scale with the incident wavelength like $\Delta \theta_0 \propto (\lambdain)^2$ and $\Delta \theta_{1,2,3} \propto (\lambdain)^3$, respectively.
Our result proves that the scattering matrix approach to QCRB~\cite{Bouchet2021} is valid for a single quantum dipole scatterer. The relative error bound of a polarizability estimate, $\Delta \chi_0 /\chi_0 \geq 1/2\sqrt{N^{\rm sc}}$, and the error bounds of position estimates relative to the wavelength, $\Delta \vr_0/\lambdain \geq \frac{1}{4\pi}(\sqrt{5},\sqrt{5/2},\sqrt{5/7})/\sqrt{N^{\rm sc}}$, are all determined by the inverse square root of the number of scattered photons, $N^{\rm sc} = \sigma_{\rm tot}\Phi$. 

While the scattering process is taking place ($|t| \sim \tau$), the QCRB improve drastically with the transient population of short-wavelength modes, i.e., enhanced near-field amplitudes around $\vr_0$. The QFI maxima at $t=0$ scale with the flux $\Phi/\tau$, independent of the temporal width or shape of the incident pulse. They also diverge for a point dipole, rendering this common idealisation invalid here.

\begin{figure}
  \begin{overpic}[width=\columnwidth]{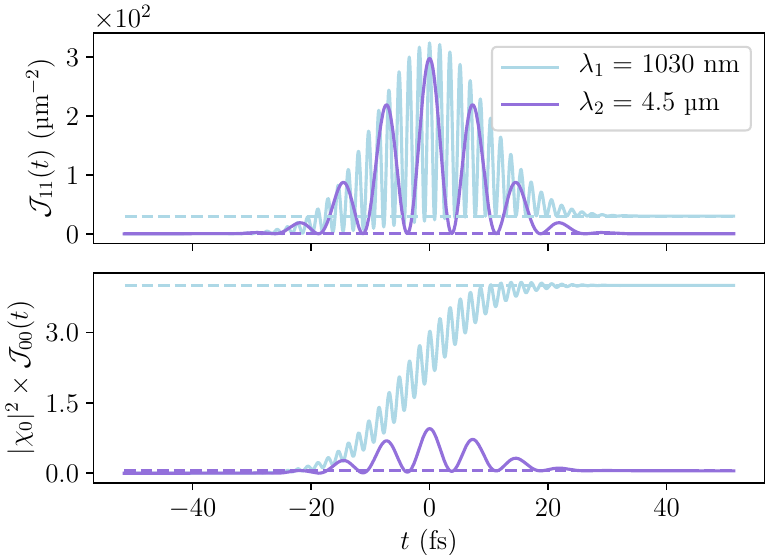}
    \put (16, 60) {\large (a)}
    \put (16, 28) {\large (b)}
  \end{overpic}
  \caption{
  \added{QFI as a function of time for estimating (a) the position $\theta_1 = x_0$ and (b) the polarizability $\theta_0 = \chi_0$  of a dipole scatterer with radius $\SI{35}{nm}$ and polarizability $\chi_0 = \SI{12.6}{nm^3}$, at two incident wavelengths $\lambda_{1,2}$. The incident photon flux at $\lambda_1$ is set such that $N^{\rm sc}=1$, the one at $\lambda_2$ is increased by $\lambda_2/\lambda_1 \approx 4.4$ for a comparable QFI peak value in (a). The dashed lines mark the far-field values; they differ by a factor $1.3\times 10^3$ in (a) and by $76$ in (b).
  These ratios differ slightly from those obtained using eq. \eqref{FarFieldQFI}, as the latter assumes a wave packet that is much longer than one wavelength.}
  }
  \label{fig:QFI}
\end{figure}

Figure~\ref{fig:QFI} shows how the QFI about (a) $x_0$-position and (b) polarizability evolves in time. We assume light pulses of central wavelengths $\lambdain_1 = \SI{1.03}{\mu m}$ and $\lambdain_2 = \SI{4.5}{\mu m}$ and
temporal width $\pulsewidth_1 = \pulsewidth_2 = \SI{24}{fs}$, corresponding to $\alpha_\keps^{\rm in} \propto E^{\rm in} e^{-(k-2\pi/\lambdain)^2(c\pulsewidth)^2/2\pi} / i\sqrt{k}$ with $\vk = k\ve_z$ and $\ve_\keps = \ve_x$. The 
scatterer has the size $a_0 \approx \SI{35}{nm}$, the polarizability $\chi_0 = \SI{13.0}{nm^3}$, and is resonant to $2\pi c/\omega_0 = \SI{100}{nm}$; see Sec.~V in \cite{Supp} for additional results. As the light pulse approaches, the information content in the field builds up and oscillates at about twice the optical frequency. The peak position information is reached when the pulse hits the scatterer around $t=0$, amplifying the far-field values here by factors of $10.8$ and $\SI{1.3e+4}{}$, respectively. The peak value grows like $(\lambdain/ a_0)^{4}$ with decreasing scatterer size $a_0 \to 0$ assuming constant polarizability. 
The oscillations in Figure~\ref{fig:QFI} (b) show that information about the polarizability is enhanced in the near-field. While the QFI never exceeds the far-field limit here, it would for smaller $a_0$, amplifying like $(\lambdain/a_0)^2$ for $a_0\to 0$. While the position uncertainty does not relate in a simple manner to the transient number of scattered photons $N^{\rm sc} (t)$ in the near field, we find that the QCRB on polarizability estimates obeys $\Delta\chi_0 /\chi_0 \geq 1/2\sqrt{N^{\rm sc}(t)}$ at all times.

\textit{Discussion.---}We derived the (Q)FI in the fields scattered by a dipole both in a phenomenological model and in a time-dependent quantum scattering model. 
The former assumes an idealized time-integrating detector that could potentially be realized in experiments. We obtain CRB for location and polarizability estimation that improve in the near field with the third and the second power of the detector distance, respectively. 

The quantum model
provides us with a snapshot of the information content in the field state at a given point in time. The QCRB 
depend on the scatterer size $a_0$, vanishing like $(a_0/\lambdain)^2$ and $a_0/\lambdain$ for location and polarizability estimation, respectively. Our calculations confirm that the transient state of the near field contains more information about the scatterer than what photodetectors could pick up at a distance. 
Information flows back and forth between the dipole and the surrounding field, causing a pronounced oscillatory enhancement of the QFI during the scattering process, $|t|\lesssim \pulsewidth$, even though a fraction of the incident pulse energy has not reached the scatterer yet. 
After the interaction ceases, the near-field information is irrevocably lost.
The far-field QCRB derived from \eqref{FarFieldQFI}, $\sqrt{N^{\rm sc}}\Delta \chi_0 /\chi_0 \geq 0.50$ and $\sqrt{N^{\rm sc}}\Delta \vr_0/\lambdain \geq (0.18,0.13,0.07)$, are independent of $a_0$ as long as $a_0 \ll \lambdain$. They provide a lower bound for microscopy applications, regardless of the light collection geometry~\cite{Dong2021FundamentalPhotometry}. 

Our analysis of the textbook example of dipole radiation touches upon foundational concepts such as ultraviolet divergences and gauge invariance. The near-field QFI diverges in the point-dipole limit, which forced us to introduce a high-frequency cutoff amounting to a finite size $a_0$ of the dipole scatterer. At the same time, the QFI depends on the chosen electromagnetic gauge that fixes the light-matter coupling Hamiltonian \cite{Stokes2022ImplicationsElectrodynamics}, 
\added{because} 
gauge transformations that depend on the dipole position $\vr_0$ can change how much information about $\vr_0$ is contained in the (transverse) field degrees of freedom. By fixing the multipolar gauge, we ensured that any information exchange between the dipole scatterer and a model detector comprised of dipoles is exclusively mediated by the transverse field (see Sec.~VIII in \cite{Supp})\added{, thus setting}
a fundamental bound on the achievable measurement precision with standard photo-detectors.

Our near-field assessment compares favorably with far-field super-resolution techniques like single-molecule localization microscopy~\cite{Rust2006Sub-diffraction-limitSTORM,Betzig2006ImagingResolution} or spatial mode demultiplexing~\cite{Tsang2016, Pushkina2021SuperresolutionField}. Our results show that, when tracking particles in the near field, one could achieve a higher signal-to-noise ratio per detected photon. This could facilitate tracking~\cite{Taylor2019InterferometricMembrane} within sensitive biological specimens~\cite{Waldchen2015Light-inducedMicroscopy} at even higher speed and precision.

Harnessing the near-field advantage comes with the experimental challenge of placing a physical detector into the near field. 
\addded{This has two consequences that must be analysed with a specific detector geometry in mind. First, the detector changes the mode structure of the electromagnetic field in its vicinity. 
Second, near-field detectors suffer from coupling inefficiencies.
For example, photon-induced near-field electron microscopy~\cite{Barwick2009Photon-inducedMicroscopy} does not affect the field mode structure, but suffers from a limited conversion efficiency between light and electrons. The two-dimensional detectors in optical near-field electron microscopy~\cite{Marchand2021OpticalMicroscopy} have a limited effect on the mode structure and reach efficiencies of a few percent. In the more common near-field scanning optical microscopy~\cite{NovotnyB.2012PrinciplesNano-optics}, a nanotip or aperture scans across the sample. It changes the local mode structure significantly, and only picks up a fraction of the near-field light.}

\addded{We have derived fundamental precision bounds for a single, short, and weak probe pulse of light that interacts with a sub-wavelength scatterer in free space. For consecutive pulses of light, the measurement back-action on the particle must be taken into account: light scattering will transfer momentum to the particle, which adds to the uncertainty of subsequent position measurements. For a single pulse, this effect can be ignored, because the measurement is finished before the induced motion will have a significant effect. In many microscopy applications, the scatterer is fixed on a cover slide and the momentum transfer thus irrelevant. In case measurement back-action does play a role, our analysis still bounds the information obtainable from each single probe pulse and thus quantifies the trade-off between gain of knowledge and back-action noise.}

\addded{Another follow-up research direction would be to specify
a detection mechanism based on, e.g., dipole-dipole interactions, which} could resolve the subtleties regarding gauge freedom. It will further be interesting to compare our 
scattering treatment to Markovian quantum trajectory models~\cite{Molmer2014}, which describe the information flow out of the scatterer
as a continuous measurement process. 
Our findings could also be extended to the radio-frequency domain, provided that an appropriate noise model is chosen. Potential applications would range from communication and positioning~\cite{Wymeersch2020AField} to the design of avalanche safety equipment~\cite{Ayuso2015ARescue}.

\begin{acknowledgments}
    We acknowledge fruitful discussions with Jonathan Dong. This project has received funding from the European Union's Horizon 2020 research and innovation programme under grant agreement No 101017902.
\end{acknowledgments}

\bibliography{references,addref}


\clearpage

\onecolumngrid
\begin{center}
\textbf{\large Supplemental Material}
\end{center}
\twocolumngrid

\setcounter{figure}{0}

\renewcommand{\thefigure}{S\arabic{figure}}
\renewcommand{\theequation}{S\arabic{equation}}

\section{Brief Introduction to Parameter Estimation Theory}\label{sec:ParameterEstimationIntro}

We provide an outline of basic concepts in parameter estimation theory and introduce the all-important Cramér-Rao precision bound. A more comprehensive introduction can be found in \cite{van2007parameter,Ly2017AInformation}.

\subparagraph{Parameter estimation from measurement data}
Estimation theory is concerned with experimental procedures producing data of measurement outcomes in order to infer one or more underlying system parameters $\vtheta$ that influence how likely the observed measurement outcomes are. To this end, one first formulates a theoretical model that predicts the likelihood $p(\vD|\vtheta)$ for observing any datum $\vD$ of outcomes at any given parameter value $\vtheta$. Secondly, one defines a point estimator $\vvtheta (\vD)$, i.e., a function that assigns an estimated parameter values to the observed data. A universally used example is to take as $\vvtheta (\vD)$ the maximum of the likelihood function with respect to the parameters, $\vvtheta(\vD) = \max_{\vtheta} p(\vD|\vtheta)$. 

Ideally, the chosen estimator should be unbiased, i.e., reproduce a given true parameter on average, $\la \vvtheta (\vD)\ra_{\vD} = \sum_{\vD} \vvtheta(\vD) p(\vD|\vtheta) \stackrel{!}{=} \vtheta$. In practice, however, strictly unbiased estimators are hard to come by, and one mostly operates with ones that become asymptotically unbiased in the large-data limit. Regardless of the chosen estimator, the local sensitivity of the experiment to small variations of the underlying parameters around a given $\vtheta = (\theta_0,\theta_1,\ldots)$ is measured in terms of the positive semidefinite Fisher information (FI) matrix,
\begin{equation}\label{eq:FI}
    \mathcal I_{j\ell}(\vtheta) = \sum_{\vD} \frac{1}{p(\vD|\vtheta)} \frac{\partial p(\vD|\vtheta)}{\partial\theta_i} \frac{\partial p(\vD|\vtheta)}{\partial\theta_j}
\end{equation}
The greater the norm or the eigenvalues of this matrix at $\vtheta$, the stronger the impact of small parameter deviations $d\vtheta$ on the likelihood of outcomes and, hence, the higher should be the achievable estimation precision in the vicinity of $\vtheta$. This can be cast into a stringent mathematical inequality, the Cram\'{e}r-Rao bound: Assume a fixed true $\vtheta$ and let $\Delta \theta_j^2 = \la [\vartheta_j(\vD) - \theta_j]^2\ra_{\vD}$ be the mean square deviation of the $j$-th parameter's estimate from the true value $\theta_j$---quantifying the measurement precision locally around $\vtheta$. For an unbiased estimator $\vvtheta$ based on large data (or many measurement repetitions), the so defined precision (per single repetition) obeys the inequality
\begin{equation}\label{eq:CRB_app}
    (\Delta\theta_j)^2 \geq [\mathcal I(\vtheta)^{-1}]_{jj}
\end{equation}
as also stated in the main text. In practice, one can often reach close to this bound.
If the data is composed of $N$ independent repetitions or identical trials or counts, the likelihood is a product of $N$ identical single-trial likelihoods and the corresponding FI is simply $N$ times the single-trial FI. The Cram\'{e}r-Rao bound thus takes shot noise into account.

\subparagraph{Ideal photo-detector surface}
In our phenomenological near-field dipole radiation model, the parameters to estimate are the polarizability and the position of the scatterer, $\vtheta = (\chi_0, x_0, y_0, z_0)$. 
For the measurement, we assume a planar or hemispherical surface comprised of individual independent photodetector pixels. That is, the measurement data is a collection of integer count numbers $n_{\vr}$ representing the outcomes of all detector pixels at positions $\vr$, with surface areas $dA$ and normal vectors $\ve_{\vn}$. As we are concerned with mathematically tractable fundamental bounds on the attainable estimation precision, we make two idealizing simplifications in our detector model, along the lines of earlier studies \cite{Ram2006,Chao2016}. Realistic detectors may not fully reach these bounds due to technical limitations.

Firstly, we follow standard practice and describe the photodetection in each pixel as a \textit{statistically independent} Poisson process integrating over the recording time $\tau$: given an average light intensity $I (\vr,\vtheta)= \ve_{\vn} \cdot \vS (\vr,\vtheta)$ that illuminates the pixel, the probability of detecting $n$ counts is $p(n|\vr,\vtheta) =  e^{-\bar{n} (\vr,\vtheta)} [\bar{n} (\vr,\vtheta)]^n /n!$, where $\bar{n} (\vr,\vtheta) = I (\vr,\vtheta) \tau d A /\hbar \omega$ denotes the average number of photons the pixel absorbs during recording time. Photon shot noise is thus accounted for.
For a pair of, say, neighboring pixels illuminated by approximately the same intensity, statistical independence means that the probability to count $n_1$ photons in the first and $n_2$ photons in the second pixel is given by the product $p(n_1|\vr,\theta) p(n_2|\vr,\theta)$. Similarly, the likelihood of the whole collection of measurement data is given by the product of the individual Poisson count distributions over all detector pixels, $p(\vD|\vtheta) = \prod_{\text{pixels}} p (n|\vr,\vtheta)$. Cross-talk between the pixels, which would only degrade the detector resolution, is not taken into account. Notice also that, rather than assuming a stationary radiation intensity, we consider narrow-band pulses with a slowly varying input amplitude $E^{\rm in} (t) = E^{\rm in} g(t)$. We can thus safely neglect the variation of the dipole fields \eqreff{scattered_field} over the pulse spectrum and simply replace the recording time by an integral over the pulse envelope, $\tau = \int d t\, |g(t)|^2$. Formally, this corresponds to modeling each pixel's count distribution by an inhomogeneous Poisson process with a time-dependent rate.

Secondly, we assume that the detector pixels are small compared to the length scale over which the intensity varies, so as to approximate the sum over pixels that appears in the FI matrix associated to the product likelihood $p(\vD|\vtheta)$ by a surface integral. Explicitly, the FI matrix of a product of independent likelihoods is the sum of the FI matrices of these likelihoods, so that
{\allowdisplaybreaks\begin{align}
    &\mathcal{I}_{j\ell} (\vtheta) = \sum_{\text{pixels}} \sum_{n=0}^\infty p(n|\vr,\vtheta) \frac{\partial \ln p(n|\vr,\vtheta)}{\partial \theta_j} \frac{\partial \ln p(n|\vr,\vtheta)}{\partial \theta_\ell} \nonumber \\
    &= \sum_{\text{pixels}} \frac{\partial \bar{n}(\vr,\vtheta) }{\partial \theta_j} \frac{\partial \bar{n}(\vr,\vtheta)}{\partial \theta_\ell} \sum_{n=0}^\infty p(n|\vr,\vtheta) \left[ \frac{n}{\bar{n}(\vr,\vtheta)} - 1 \right]^2 \nonumber \\
    &= \sum_{\text{pixels}} \frac{1}{\bar{n}(\vr,\vtheta)}\frac{\partial \bar{n}(\vr,\vtheta) }{\partial \theta_j} \frac{\partial \bar{n}(\vr,\vtheta)}{\partial \theta_\ell} \nonumber \\
    &\approx \frac{\tau}{\hbar \omega}\int_{\text{pixels}} \frac{dA}{I(\vr,\vtheta)} \frac{\partial I(\vr,\vtheta) }{\partial \theta_j} \frac{\partial I(\vr,\vtheta)}{\partial \theta_\ell},
\end{align}}
as stated in Eq.~\eqreff{eq:FIMPoissonContinuum} in the main text. Here we have used that $\bar{n}(\vr,\vtheta)$ gives both the mean and the variance of each Poisson distribution.

\subparagraph{Quantum bound on parameter estimation}
Suppose the system under observation at a given point in time $t$ is described by a quantum state $\varrho (t,\vtheta)$, which depends on the parameters $\vtheta$ we seek to infer. A measurement protocol with we conduct on this state is generally described by a POVM $\mathcal{M} = \{\oM_{\vD}\}$: a set of positive semidefinite operators $\oM_{\vD}$ associated to the measurement outcomes $\vD$ that obey $\sum_{\vD} \oM_{\vD} = \id$, such that the likelihood for obtaining outcome $\vD$ is $P^{\mathcal{M}}(\vD|\vtheta) = \tr [ \oM_{\vD} \varrho(t,\vtheta)]$. The associated FI matrix $\mathcal{I}^{\mathcal{M}} (\vtheta,t)$ determines the estimation precision achievable in this protocol via the CRB \eqref{eq:CRB_app}---but a different POVM may yield a better precision, i.e., a FI matrix with greater eigenvalues. 
In order to place a fundamental bound on the physically attainable precision given the quantum state $\varrho (t,\vtheta)$, we must 'optimize' the FI matrix over all possible measurements. 

The optimization is straightforward in the case of a single parameter $\theta$. The FI is then a scalar quantity, and we can take the maximum over all POVMs: $\mathcal{J} (\theta,t) := \max_{\mathcal{M}} \mathcal{I}^{\mathcal{M}} (\theta,t)$, also known as the quantum Fisher information (QFI). It sets the fundamental quantum Cram\'{e}r-Rao bound (QCRB) on estimation precision, $\Delta \theta^2 \geq \mathcal{J}^{-1} (\theta,t)$. The QFI is uniquely determined by the quantum state; it can be expressed as $\mathcal{J} (\theta,t) = \tr [\varrho (t,\theta) \oL^2 ]$, in terms of the so-called symmetric logarithmic derivative operator, $\oL = \oL^\da$, defined implicitly through the Lyapunov equation $\partial_{\theta}\varrho(t,\theta) = \{ \oL, \varrho(t,\theta)\}/2$. 

A natural extension to multi-parameter estimation problems is the QFI matrix
\begin{equation}\label{appQFI}
    \mathcal J_{j\ell}(\vtheta,t) = \frac{1}{2}\tr[\varrho(t,\vtheta)\{\oL_j,\oL_\ell\}] \geq 0,
\end{equation}
with the symmetric logarithmic derivative operators defined through
\begin{equation}\label{appSLD}
    \frac{\partial}{\partial\theta_j} \varrho(t,\vtheta) = \frac{1}{2} \{ \oL_j, \varrho(t,\vtheta) \}.
\end{equation}
The QFI matrix upper-bounds the FI matrix associated to any measurement, $\mathcal{J}(\vtheta,t) \geq \mathcal{I}^{\mathcal{M}} (\vtheta,t)$, in the usual sense that $\mathcal{J} - \mathcal{I}^{\mathcal{M}}$ has non-negative eigenvalues. 
From this follows the QCRB stated in the main text, $\Delta \theta_\ell^2 \geq [\mathcal{J}^{-1}(\vtheta,t)]_{\ell \ell}$.

\subparagraph{QFI of electromagnetic field states}

The operators $\oL_j$ defined implicitly in \eqref{appSLD} are difficult to compute in practice for general quantum states. However, in the case of bosonic Gaussian states, which are fully determined by their first and second moments in the bosonic field operators, there are closed-form expressions for them and the QFI \cite{Safranek2017}. We make use of these expressions in Supplementary Section \ref{sec:QFI_Appendix}, to calculate the approximate QFI stated in Eq.~\eqref{GaussianQFI} in the main text. 
We remark that, unlike the FI that we evaluate for a time-integrating detector, the QFI represents the information about the scatterer contained in the field state at a given time $t$ in a given gauge (see Supplementary Section \ref{sec:GaugeInvariance}). Our results show that, at times $t$ when the scattering process of the input pulse at the dipole scatterer is taking place, the QFI exhibits a near-field enhancement, despite the fact that part of the input pulse has not reached the scatterer yet. A fair comparison between the classical FI in our phenomenological model and the QFI can be done in the far field: once the scattering process is completed, the acquired information is stored in the phases and amplitudes of asymptotically outgoing plane waves. These propagate freely and preserve their information content \cite{hüpfl2023continuity}, which one can, in principle, read out with help of phase-sensitive, direction-resolving detectors.


    \newpage

\section{CRB for a Classical Hertz Dipole}
\label{app:ClassicalDipole}

Here we complement the phenomenological approach of the main text, assuming stationary radiation from a classical field-induced Hertz dipole. We provide additional results for planar detectors, a hemispherical detector, and we discuss the behavior of the CRB in the near field. Finally, we state the scattering fields for a regularized finite-size dipole instead of a point dipole, for comparison to the quantum model. 

\begin{figure}

	\begin{minipage}{.45\textwidth}
		\includegraphics[width=\linewidth]{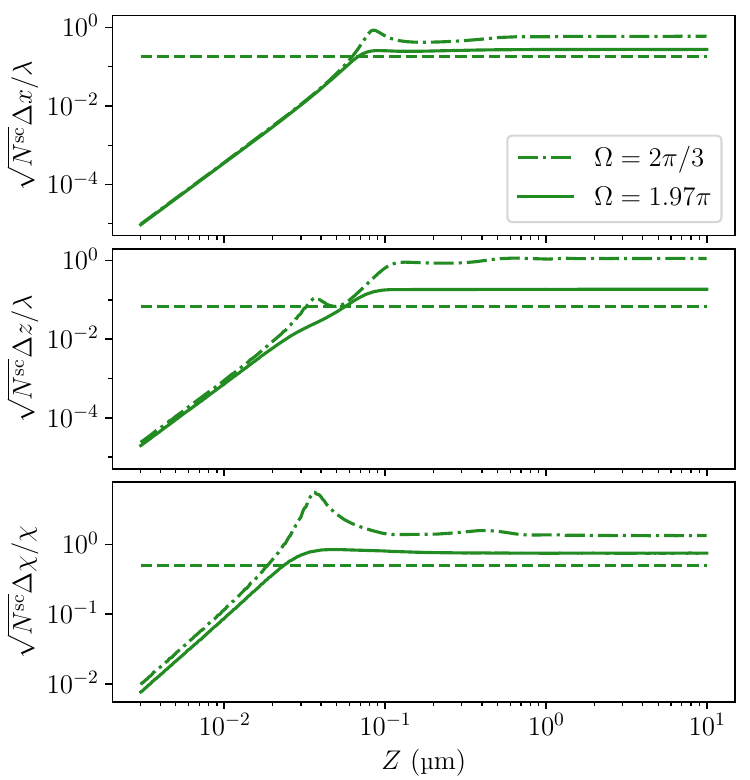}
	    \caption{
	    CRBs for two square-shaped planar detectors of different sizes as a function of the distance to the sample. The detector size is given in terms of the covered solid angle $\Omega$. The sample is a point dipole ($a_0=0$) of polarizability $\chi_0 = \SI{12.6}{nm^3}$, the incident wavelength is $\lambdain=\SI{532}{nm}$. As expected, the detector covering a larger solid angle (solid line, same as in Fig.~\reff{fig:CRB_plots_planar} of the main text) captures more information. The difference is less pronounced for $\Delta x$, indicating more information originating from pixels close to the optical axis.
	    }
	    \label{fig:test12}
		\includegraphics[width=\linewidth]{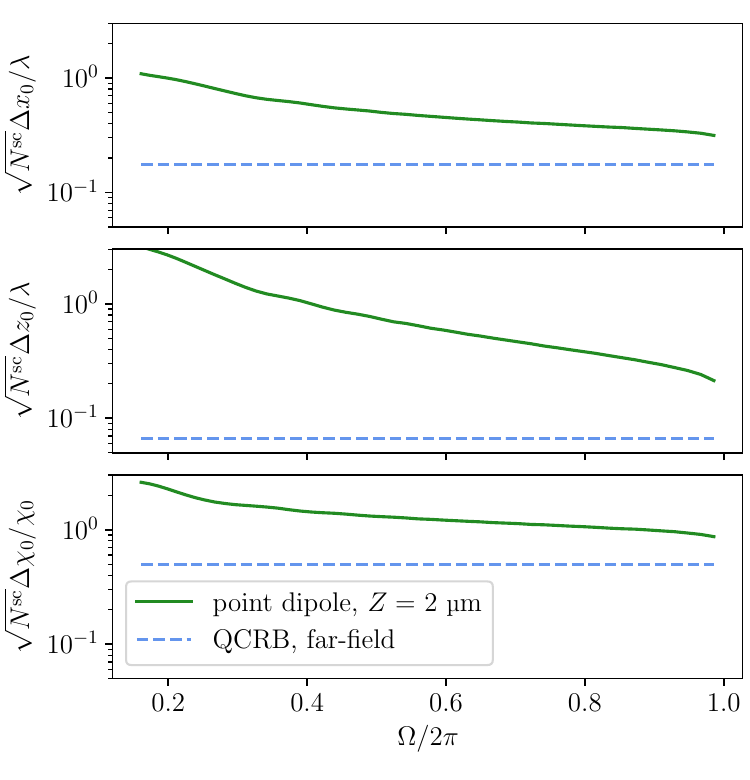}
	    \caption{
	    CRBs for variable detector size. Each detector is placed $\SI{2}{\mu m}$ in front of the sample, square-shaped and planar, parallel to the $xy$-plane, and subtends a solid angle $\Omega$. The polarizability is $\chi_0 = \SI{12.6}{nm^3}$, and the point dipole approximation is used ($a_0=0$).
	    The incident wavelength is $\lambdain=\SI{1.03}{\mu m}$.
	    }
	    \label{fig:test16}
	\end{minipage}
	
\end{figure}

\subparagraph{Planar detector}
In the main text, we have assumed a fairly large planar detector covering almost the entire range of scattering angles from $-\pi/2$ to $\pi/2$ into the forward (or backward) half-space. Namely, for the Cram\'{e}r-Rao bounds plotted in Fig.~\reff{fig:CRB_plots_planar}, we have integrated over angles up to $\pm 0.495\pi$. Such a detector covers a solid angle of $1.97\pi$ measured from the origin. In practice, one can capture a major part of the information already in much smaller detectors. In Fig.~\ref{fig:test12}, we compare the CRBs in forward direction from the main text (solid lines) to the CRBs evaluated for a planar detector covering a smaller solid angle (dash-dotted). The near-field behaviour, in particular, hardly differs. Figure \ref{fig:test16} shows how the CRBs scale with the detector size (given in terms of the covered solid angle $\Omega$), at a fixed far-field distance $Z=\SI{2}{\mu m}$.

\begin{figure}
	\begin{overpic}[width=\columnwidth]{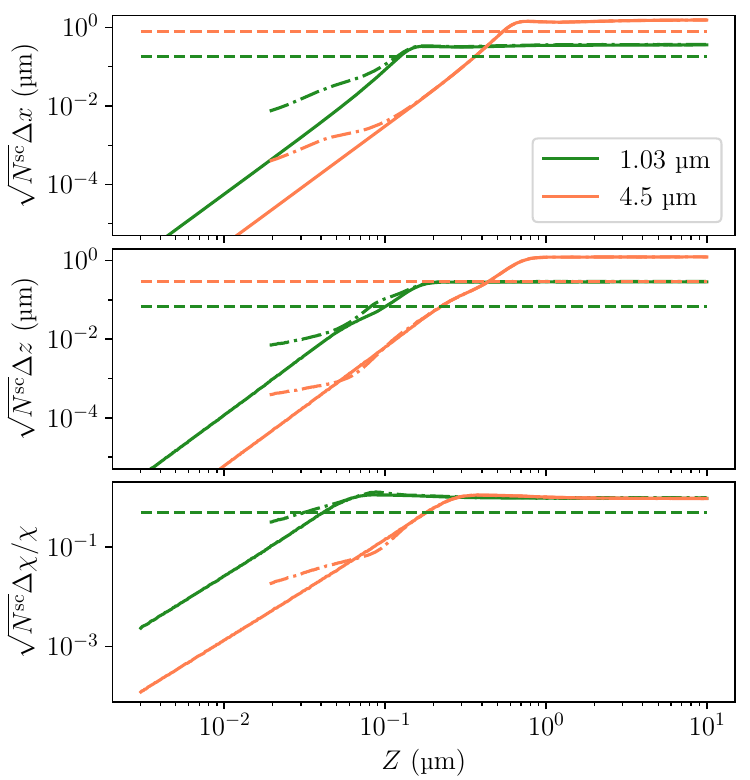}
	  \put (18, 86) {\large (a)}
	  \put (18, 54) {\large (b)}
	  \put (18, 25) {\large (c)}
	\end{overpic}
	\caption{
	Comparison of normalized Cramér-Rao bounds for two incident wavelengths. Solid lines: Cramér-Rao bounds for planar detector covering a solid angle of $1.86\pi$ placed a distance $Z$ in front of a point dipole scatterer. Dash-dot lines: Cramér-Rao bounds for the same detector with a sample of radius $a_0=\SI{35}{nm}$.
	Dashed lines: Quantum Cramér-Rao bounds in the far-field. 
	}
	\label{fig:test8}
\end{figure}

Finally, in Fig.~\ref{fig:test8}, we compare the CRBs for two incident infrared wavelengths at the same number of scatterer photons and at a given scatterer in the point dipole limit (solid) or with finite size (dash-dotted, $a_0=\SI{35}{nm}$). To this end, the results for (a) $x$- and (b) $z$-estimation are \textit{not} normalized to the wavelength, but given in absolute units. We observe that the near-field improvement is more pronounced at the greater wavelength.

\subparagraph{Hemispherical Detector}

\begin{figure}[h]
    \centering
    \includegraphics[width=0.73\linewidth]{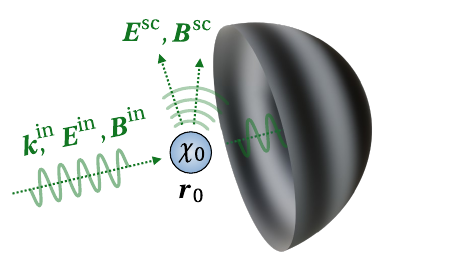}
    \caption{Sketch of the hemispherical detector setup. The detector surface is a hemisphere of radius $R$ with the scattering dipole at its origin and its pole pointing in the direction of the $\vk$-vector of the incident light field ($\vk^{\rm in}, \boldsymbol E^{\rm in}$, $\boldsymbol B^{\rm in}$). The scattered light is denoted by ($\boldsymbol E^{\rm sc}$, $\boldsymbol B^{\rm sc}$).}
    \label{fig:sketch_spherical}
\end{figure}

As an alternative to the planar detector discussed in the main text 
(Fig.~\reff{fig:sketch}), 
we consider a hemispherical detector of radius $R$ around the dipole scatterer, oriented in forward ($Z>0$) or backward ($Z<0$) direction of the incident light, as sketched in Fig.~\ref{fig:sketch_spherical}. The task is to resolve small deviations of the scatterer position and polarizability based on the detected photons from the scattered light.

\begin{figure*}
\includegraphics[width=\linewidth]{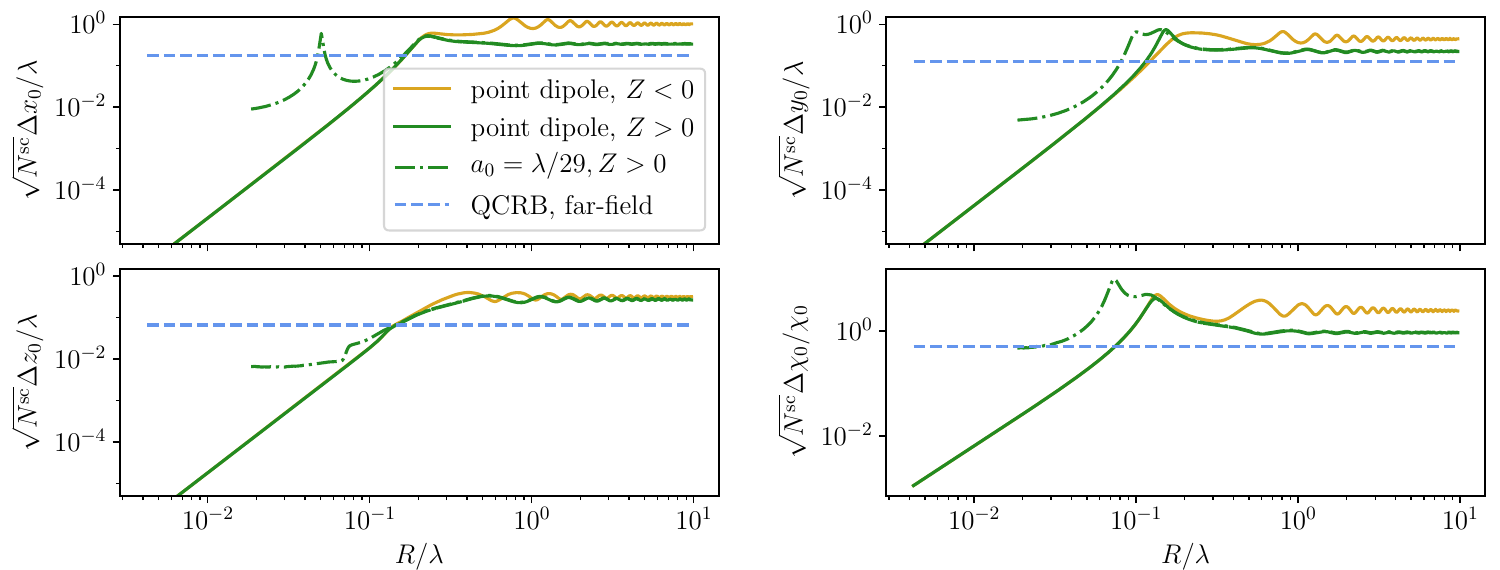}
\caption{Cram\'{e}r-Rao bounds (hemispherical detector) and far-field quantum CRB, normalized by the total number of scattered photons, for $\chi_0 = \SI{13.0}{nm^3}$ at $\lambdain = \SI{1.03}{\mu m}$. The detector is slightly smaller than a full hemisphere and covers a solid angle of $1.84\pi$. Note that the signal stems from the interference of the scattered wave and the unscattered plane wave. Their relative phase depends on the distance to the scatterer, leading to the oscillations in the far field. We further note that, with linearly polarized excitation light, the CRB for the estimation of the x and y position of the scatterer differ slightly. The orange line corresponds to the hemisphere oriented in backward direction ($Z<0$). 
}
\label{fig:CRB_plots_spherical}
\end{figure*}

Using the same detector model and parameters as for Fig.~\reff{fig:CRB_plots_planar} 
in the main text, we plot the CRBs on position and polarizability in Fig.~\ref{fig:CRB_plots_spherical}. The results are qualitatively similar to those of the planar detector, except for small oscillations as a function of the radius $R$ due to interference effects at the detector edge.

\subparagraph{Asymptotic Behavior}

Near the dipole scatterer ($\kin r \lesssim 1$), the scattered fields \eqreff{scattered_field} in the main text scale like $\vE^{\rm sc} \sim 1/r^3, \vB^{\rm sc} \sim 1/r^2$. Since we can also assume that $(4\pi \chi_0)^{1/3} \ll |Z|,R$ for most of the plotted range of detector distances, the terms in the Poynting vector,
\begin{equation}\label{Poynting_vector_four_terms}
  \vS (\vr) = \frac{ \text{Re}\{[\vE^{\rm in} (\vr) +\vE^{\rm sc} (\vr)] \times [\vB^{\rm in} (\vr) +\vB^{\rm sc} (\vr)]^*\}}{2\mu_0},
\end{equation}
obey the hierarchy $   \vE^{\rm in}\times\vB^{\rm in} \gg
   \vE^{\rm sc}\times\vB^{\rm in}
   + \vE^{\rm in}\times\vB^{\rm sc} \gg
   \vE^{\rm sc}\times\vB^{\rm sc}$.
Therefore, in the formula \eqreff{eq:FIMPoissonContinuum} for the FI matrix $\mathcal I$ in the main text, the intensity $I$ is dominated by the incident light term, while the derivatives of the intensity with respect to the scatterer parameters are dominated by the cross terms. To leading order in $\kin r$ with $r\sim |Z|,R$, we have $\partial I/\partial\theta_0 \sim 1/r^3$ and $\partial I/\partial\theta_{j>0} \sim 1/r^4$. 
Integration over the detector surface contributes another factor $2\pi R^2$ in the hemispherical case. In the planar case, the relevant detector area in the near field is of the order of $\pi Z^2$. Hence, the diagonal entries of the FI matrix scale like $\mathcal{I}_{00} \sim R^{-4},Z^{-4}$ and $\mathcal{I}_{jj} \sim R^{-6},Z^{-6}$ for $j=1,2,3$. Accordingly, the CRB for polarizability and position estimates scale like $\Delta \chi_0 \sim R^2,Z^2$ and $\Delta \vr_0 \sim R^3,Z^3$, respectively, matching the slopes in Fig.~\reff{fig:CRB_plots_planar} in the main text and in Fig.~\ref{fig:CRB_plots_spherical}.

In the far field, we simply have $\partial I/\partial\theta_j \sim R^{-1},|Z|^{-1}$ for all $j$, whereas $I \to const$. Hence, the entries of the FI matrix should approach a constant value for $R,|Z| \to \infty$, which is also in accordance with our results.

\begin{figure*}
\includegraphics[width=\linewidth]{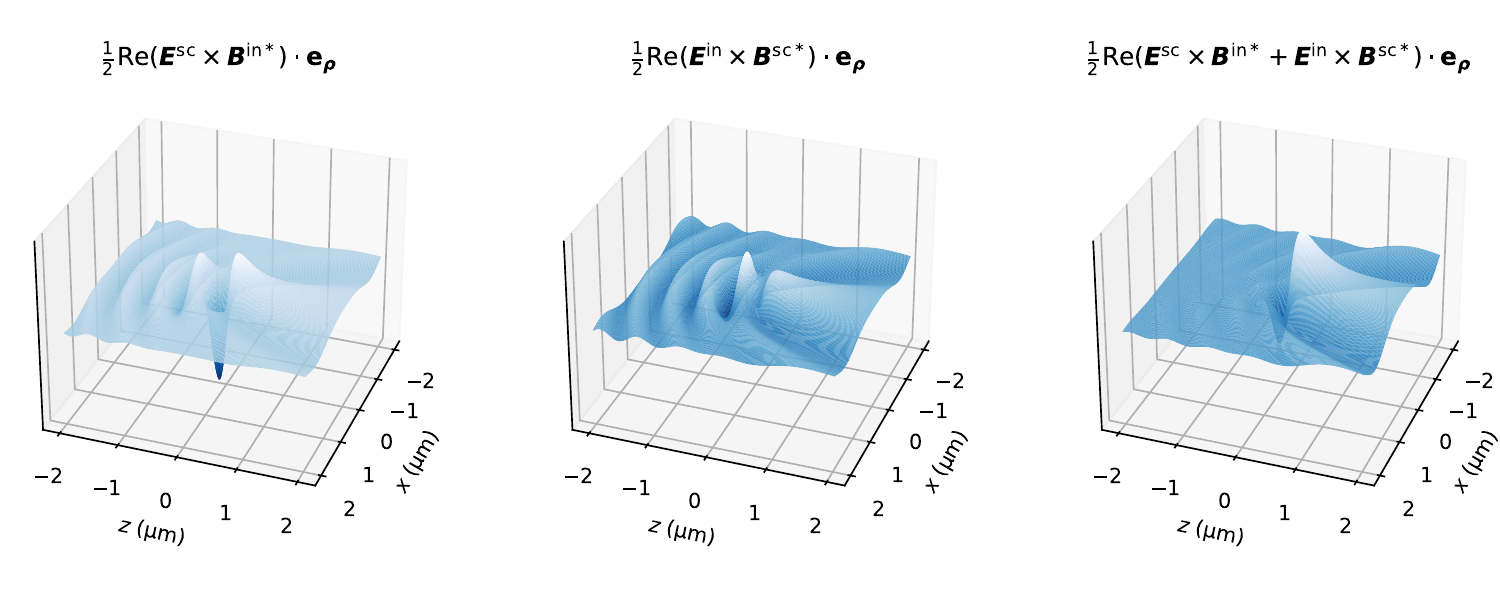}
\caption{ 
Interference terms between background and scattered field plotted over a $\SI{4}{\mu m} \times \SI{4}{\mu m}$ rectangular area in the plane defined by $y = 5 a_0$ with the scatterer radius $a_0 = \SI{35}{nm}$. This value for $y$ was chosen to be close to the scatterer but sufficiently far away to avoid large near-field values. The two interference terms (1st and 2nd panel) are the main contributions to the time-averaged Poynting vector for weak coupling. The rightmost plot shows the sum of the two terms. They interfere largely destructively for $z<0$ and constructively for $z>0$, because the electric and magnetic fields are even and odd functions under $\vrho \to -\vrho$, respectively.
The ripples in the $Z<0$ half-space are attributable to the path length difference between the forward-traveling background light and the backward-traveling scattered light.
}
\label{fig:interference_terms}
\end{figure*}

In Fig.~\ref{fig:CRB_plots_spherical} and in Fig.~\ref{fig:CRB_plots_planar} of the main text, we saw that the optimal estimation precision is worse for a detector placed behind the sample ($Z<0$) for all parameters except $z_0$. The reason is destructive interference between the incident and the scattered field, as can be seen in Fig. \ref{fig:interference_terms}. Therein, we plot the expressions $\frac{1}{2}\text{Re}[\vE^{\rm sc}(\vr) \times \vB^{\rm in *}(\vr)] \cdot \ve_{\vrho}$ and $\frac{1}{2}\text{Re}[\vE^{\rm in}(\vr) \times \vB^{\rm sc *}(\vr)] \cdot \ve_{\vrho}$ as well as their sum, over the plane $y = \SI{175}{nm}$.
At weak coupling, these terms are the main contributors to the relevant measurement  signal and the FI. The rightmost plot shows that the two terms interfere mainly destructively for $Z<0$ and mainly constructively for $Z>0$. This is consistent with the fact that $\vE^{\rm sc}(\vr)$ and $\vB^{\rm sc}$ are even and odd functions of $\vrho$ respectively. When estimating $z_0$, this effect is compensated by the path length difference between background radiation and light scattered back into a $Z<0$ detector, which manifests as the ripples at negative $z$-values in Fig. \ref{fig:interference_terms}.

\begin{widetext}
\subparagraph{Finite-size scatterer}
For the case of a regularized dipole scatterer of effective size $a_0>0$, we interpret the field expectation values \eqref{eq:Esc_regularized} and \eqref{eq:Bsc_regularized} obtained from the quantum model in Supplementary Section \ref{app:FieldExpectationValues} as classical fields with $\chi_0 \equiv \chi(c\kin)$. 
Explicitly, 
\begin{align}
  \vE^{\rm sc }(\vr) &= \frac{(\kin)^3 \chi_0 E^{\rm in}}{2\pi i} \left[
  \frac{(\ve_{\vrho} \times \ve_x) \times \ve_{\vrho}}{\kin\rho}\, \mathcal E_1(\rho)
  \right. 
  + \left. \left(
  \frac{\mathcal E_2(\rho)}{(\kin\rho)^2} - \frac{\mathcal E_3(\rho)}{(\kin\rho)^3}
  \right) (\ve_x - 3\ve_{\vrho}(\ve_x\cdot\ve_{\vrho}))
  \right], \nonumber
  \\
  \vB^{\rm sc }(\vr) &= \frac{(\kin)^3 \chi_0 E^{\rm in}}{2\pi c i} \left[
  \frac{\mathcal E_2(\rho)}{i\kin\rho} - \frac{\mathcal E_3(\rho)}{i(\kin\rho)^2}
  \right] \ve_{\vrho} \times \ve_x,
\end{align}
where 
\begin{equation}\begin{aligned}
    \mathcal E_1(\rho) &= \frac{16i}{[ 4+(a_0\kin)^2]^2} \left[ \left(\frac{4\rho}{a_0(a_0 \kin)^2} + \frac{\rho-a_0}{a_0} \right)e^{-2\rho/a_0} 
    + e^{i\kin\rho} \right],
    \\
    \mathcal E_2(\rho) &= -\frac{16}{[4+(a_0\kin)^2]^2} \left[i \left(\frac{ \kin(a_0-2\rho)a_0}{4} - \frac{a_0 + 2\rho}{a_0^2\kin} \right)e^{-2\rho/a_0} 
    + e^{i\kin\rho}\right],
    \\
    \mathcal E_3(\rho) &= \frac{16i}{[4+(a_0\kin)^2]^2} \left[ \left(\frac{\rho}{a_0} + 1 + \frac{(a_0 \kin)^2\rho}{4a_0}\right) e^{-2\rho/a_0} - e^{i\kin\rho} \right] .   
\end{aligned}\end{equation}
\new{In Supplementary Section \ref{app:FieldExpectationValues}, we show that these expressions are obtained by averaging the scattering fields of an ideal point dipole ($a_0=0$) over the radial dipole polarization density $\dens(\vr) = e^{-2r/a_0}/(\pi a_0^3)$ with mean radius $3a_0/2$, assuming $a_0\kin \ll 1$ and $a_0/\rho \ll 1$.}
\end{widetext}


\clearpage 
\twocolumngrid

\section{Interaction Hamiltonian}
\label{sec:InteractionHamiltonian}

Here, we review the steps leading to the dipole Hamiltonian \eqreff{interactionHamiltonian} in the main text, which describes the scatterer-field interaction.
Our starting point is the minimal coupling Hamiltonian of non-relativistic quantum electrodynamics between a single bound charge $q$ and the electromagnetic field in the multipolar (PZW) gauge with respect to the dipole position $\vr_0$~\cite{Stokes2022ImplicationsElectrodynamics},
\begin{align}\label{eq:TotalH_Multipolar}
    &\hat H = \frac{1}{2m}[\hatvp_e - q\hatvA(\hat \vr_e+\vr_0)]^2 + U(\hat \vr_e) + V_{\rm self} \\
    &+ \frac{\epsilon_0}{2}\int d^3 x \left\{ \left[\hatvPi (\vx) + \frac{1}{\epsilon_0} \boldsymbol{\hat P}_T (\vx) \right]^2 + c^2[\nabla \times \hatvA_T (\vx)]^2 \right\}. \nonumber 
\end{align}
The charge is a quantum particle trapped in a potential $U$ sourced by the opposite charge $-q$ fixed at $\vr_0$. The displacement of $q$ from $\vr_0$ is described by conjugated position and momentum operators $\ovr_e$, $\ovp_e$. The chosen gauge leads to the (infinite) Coulomb self-energy terms $V_{\rm self}$ of both charges, while the quantized light field is described by the transverse vector potential $\hatvA_T(\vx)$ and its canonical conjugate $\hatvPi (\vx)$. The full multipolar-gauge vector potential is then given by
its real-space representation
\begin{align}\label{eq:TotalA_Multipolar}
  \hatvA(\vx) &= \hatvA_T(\vx) + \nabla \int \frac{d^3k' }{(2\pi)^3} \boldsymbol{g}_T(\vk',\vx)\cdot\hatvA(\vk') \nonumber \\
  &= \hatvA_T(\vx) - \int \frac{d^3k' }{(2\pi)^3} \hatvA_T(\vk')  e^{-i\vk'\cdot\vr_0}
\end{align}
where $\boldsymbol{g}_T$ is defined as \begin{equation}\label{eq:gT}
  \boldsymbol{g}_T(\vk,\vx) = - e^{-i\vk\cdot\vr_0}\sum_\eps \ve_\keps(\ve_\keps\cdot\vx)
\end{equation}
The Hamiltonian \eqref{eq:TotalH_Multipolar} also contains the transverse polarization $\hatvP_T$ whose definition is also gauge-dependent. In the PZW gauge, it is most conveniently representable in reciprocal space:
\begin{align}\label{eq:TransvP_Multipolar}
  \hatvP_T(\vk) &= - q  [ \vg_T (\vk, \vr_0 + \hatvr_e) - \vg_T (\vk, \vr_0) ]
  \nonumber \\
  &= qe^{-i\vk\cdot\vr_0}\sum_\eps \ve_\keps (\ve_\keps\cdot\hatvr_e).
\end{align}
We work in the electric dipole, or long-wavelength approximation. This amounts to assuming that the wavelengths impinging on the scatterer are much longer than the extent of the scatterer. More concretely, we assume $\vk\cdot\hatvr_e \ll 1$. 
Fourier-transforming \eqref{eq:TotalA_Multipolar}, we have \begin{equation}
  \hatvA(\vx) = \int \frac{d^3k }{(2\pi)^3} \hatvA_T(\vk)  [e^{-i\vk\cdot\vx} - e^{-i\vk\cdot\vr_0}].
\end{equation} Substituting $\vx = \vr_0 + \hatvr_e$, and $e^{-i\vk\cdot\hatvr_e}-1 \approx 0$, this immediately yields $\hatvA(\vr_0+\hatvr_e) \approx 0$. The Hamiltonian \eqref{eq:TotalH_Multipolar} then becomes
\begin{align}\label{eq:TotalH_Multipolar2}
    \hat H &= \frac{\hatvp_e^2 }{2m}+ U(\hat \vr_e) + V_{\rm self} \nonumber \\
    &+ \frac{\epsilon_0}{2}\int d^3 x \left\{ \hatvPi (\vx)^2 + c^2[\nabla \times \hatvA_T (\vx)]^2 \right\} \\
    &+ \frac{1}{2}\int d^3 x \left\{ \frac{1}{\epsilon_0}\hatvP_T (\vx)^2 + 2\hatvP_T(\vx)\cdot\hatvPi(\vx) \right\}. \nonumber
\end{align}
The $\hatvP_T(\vx)^2$ term only involves the $\hatvr_e$ operator, and can be subsumed into $U(\hatvr_e)$ by defining $U'(\hatvr_e) = U(\hatvr_e) + \int d^3 x \hatvP_T (\vx)^2/2\epsilon_0$. The last term describes the charge-field interaction, 
\begin{align}\label{eq:HI_app}
    \hat H_I &= \int d^3 x\ \hatvPi (\vx) \cdot \boldsymbol{\hat P}_T (\vx) \\
    &= \int \frac{d^3k}{(2\pi)^3} \hatvPi (\vk) \cdot \boldsymbol{\hat P}_T^\dag (\vk)
    = q \int \frac{d^3k}{(2\pi)^3} e^{i\vk\cdot\vr_0} \hatvPi (\vk) \cdot \hatvr_e \nonumber \\
    &= q \hatvPi(\vr_0)\cdot\hatvr_e,  \nonumber
\end{align}
where we have used \eqref{eq:TransvP_Multipolar} along with the fact that $\hatvPi$ is transverse so that $\sum_{\eps=1,2}\ve_\keps(\hatvPi(\vk)\cdot\ve_\keps) = \hatvPi(\vk)$. This is the only term that entangles the field with the charge, whereas all the other terms act on either the field or the charge. Correspondingly, we define the free Hamiltonian $\hat H_0 = \hat H - \hat H_I$.

The multipolar gauge can be modified by introducing a high-frequency cutoff, $k \lesssim 1/a_0$, to avoid the high-energy divergences inherent in the dipole approximation~\cite{Stokes2022ImplicationsElectrodynamics}. This amounts to setting 
\begin{equation}\label{eq:gT_Regularized}
  \boldsymbol{g}_T(\vk,\vx) =  - \frac{e^{-i\vk\cdot\vr_0}}{[1+\tfrac{1}{4}(a_0k)^2]^2}\sum_\eps \ve_\keps(\ve_\keps\cdot\vx),
\end{equation} 
ensuring that $\vg_T(\vk,\vx) \to 0$ for $k\to\infty$.
The vector potential and polarization in \eqref{eq:TotalA_Multipolar} and \eqref{eq:TransvP_Multipolar} are then 
\begin{align}\label{eq:RegularizedGaugeTransf}
  \hatvA(\vx) &= \hatvA_T(\vx) - \int \frac{d^3k' }{(2\pi)^3} \frac{\hatvA_T(\vk')  e^{-i\vk'\cdot\vr_0}}{[1+\tfrac{1}{4}(a_0k')^2]^2}, \\
  \hatvP_T(\vk) &= \frac{qe^{-i\vk\cdot\vr_0}}{[1+\tfrac{1}{4}(a_0k)^2]^2} \sum_\eps \ve_\keps (\ve_\keps\cdot\hatvr_e). \label{eq:TransvP_Regularized}
\end{align}
The assumption $\hatvA(\vr_0 + \hatvr_e) \approx 0$ continues to hold as long as all modes with $k \geq 1/a_0$ are unpopulated. This is true, and \eqref{eq:TotalH_Multipolar2} remains valid, if the scatterer is larger than $a_0$.

We now quantize the field in the usual manner, by introducing \emph{discrete} plane-wave modes in a box of volume $L^3$ and their operators, $[\oa_\keps , \oa_\pmu^\da] = \delta_{\vk\vp}\delta_{\eps \mu}$, such that
\begin{align}\label{eq:fieldquantization}
 \hatvA_T (\vx) &= \sum_{\keps} \sqrt{\frac{\hbar}{2\epsilon_0 L^3 c k}} \ve_\keps \left( e^{i\vk\cdot\vx} \oa_\keps + e^{-i\vk\cdot\vx} \oa_\keps^\da \right), \nonumber \\
 \hatvPi (\vx) &= -i \sum_{\keps} \sqrt{\frac{\hbar ck}{2\epsilon_0 L^3}} \ve_\keps \left( e^{i\vk\cdot\vx} \oa_\keps - e^{-i\vk\cdot\vx} \oa_\keps^\da \right).
\end{align}
The continuum limit $\sum_\keps \to (L/2\pi)^3 \sum_\eps \int d^3 k$ will be carried out later.
The interaction Hamiltonian becomes 
\begin{align}\label{eq:HI_app_regularized}
    \hat H_I &= \int d^3 x\ \hatvPi (\vx) \cdot \boldsymbol{\hat P}_T (\vx)   \\
    &= \int \frac{d^3k}{(2\pi)^3} \hatvPi (\vk) \cdot \boldsymbol{\hat P}_T^\dag (\vk) \nonumber \\
    &= q \int \frac{d^3k}{(2\pi)^3} \frac{
    e^{i\vk\cdot\vr_0} \hatvPi (\vk) \cdot \hatvr_e }{[1+\tfrac{1}{4}(a_0k)^2]^2}
    \nonumber \\
    &= q \sum_\keps \sqrt{\frac{\hbar c k}{2\epsilon_0 L^3}} \frac{\ve_\keps \cdot\hatvr_e}{[1+\tfrac{1}{4}(a_0k)^2]^2} \left[
    \frac{ e^{i\vk\vr_0} }{ i } \hat a_\keps + h.c.
    \right]. \nonumber
\end{align}
Finally, we restrict the motion of the bound charge to a one-dimensional harmonic motion in $x$-direction, setting the dipole operator $q\ovr_e \equiv d_0 \ve_x (\ob + \ob^\da)$, with $\ob$ the associated ladder operator. Defining 
\begin{equation}
    \dens_{\vk} = \frac{1}{[1+\tfrac{1}{4}(a_0k)^2]^2}
\end{equation}
then leaves us with the model Hamiltonian \eqreff{interactionHamiltonian} in the main text.

By fixing the regularized gauge in \eqref{eq:RegularizedGaugeTransf} and subsequently assuming $\hatvA(\vr_0+\hatvr_e)\approx 0$, we have prevented the interaction Hamiltonian \eqref{eq:HI_app_regularized} from coupling $\hatvr_e$ to any modes with wavelength $\lambda \ll 2\pi/a_0$. Effectively, the regularizing factor in $\vk$-space relaxes the point-dipole assumption and gives the scatterer a finite transverse polarization density with profile $e^{-2r_0/a_0}/(\pi a_0^3)$. (For a derivation of this see Supplementary Section \ref{app:FieldExpectationValues}.) This is no violation of gauge invariance: mode populations are not directly measurable. What must be gauge-invariant are the probabilities of photon detection events. To calculate the latter, one must specify a concrete coupling between the detector and the system. We discuss this point in more detail later, in Supplementary Section~\ref{sec:GaugeInvariance}.

\clearpage 
\onecolumngrid 

\section{Time Evolution}
\label{sec:TimeEvol}

Here we solve the combined quantum time evolution of the harmonic field and scatterer degrees of freedom under the Hamiltonian $\oH = \oH_0 + \oH_I$, assuming an asymptotically free incident coherent pulse and the scatterer in the ground state at initial time $t_0\to -\infty$, 
$|\psi^{\rm in} (t_0) \ra =  \bigotimes_{\vk,\eps} \oD_\keps \left(\alpha^{\rm in}_{\vk\eps} e^{-ick t_0} \right) |\vac \ra \otimes |0\ra$,
with $\oD (\alpha)$ the displacement operator.
The goal of the following calculation is to integrate the Heisenberg equations of motion for the mode operators. From this, we obtain the coherent amplitudes $\valpha$ and the covariance matrix blocks $\Xi,\Upsilon$ used to calculate the quantum Fisher information matrix in the main text. 
The initial condition for the incident light amplitudes, $\alpha^{\rm in}_\keps$, is chosen such that the wave packet of the pulse is centered around the dipole position $\vr_0$ at $t=0$. 
 
In the Hamiltonian, we identify the bare terms, $\oH_0 = \sum_{\vk,\eps} \hbar c k \oa^\da_{\vk\eps} \oa_{\vk\eps} + \hbar\omega_0 \ob^\da \ob$, and the interaction Hamiltonian, 
\begin{equation}\label{eq:HI_adiab}
    \oH_{I} = (\ob + \ob^\da) d_0 \ve_x \cdot \hatvPi(\vr_0) = (\ob + \ob^\da) \sum_\keps \sqrt{\frac{\hbar c k}{2\epsilon_0 L^3}} d_0 (\ve_\keps \cdot \ve_x) \dens_{\vk} \left[ \frac{e^{i\vk\cdot\vr_0}}{i} \oa_\keps + h.c. \right].
\end{equation}
The mode operators satisfy $[\oH_0, \oa_\keps] = -\hbar ck a_\keps$ and $[\oH_0, \ob] = -\hbar \omega_0 \ob$ with respect to the bare term, and so the equations of motion for the respective Heisenberg-picture mode operators take the form
\begin{equation}\label{eq:HeisenbergEOM2}
    \frac{d}{dt} \ob_H (t) = - i\omega_0 \ob_H(t) + \frac{i}{\hbar} \left[ \oH_I, \ob \right]_H (t); \qquad \frac{d}{dt} \oa_{\keps,H} (t) =  - i c k \oa_{\keps, H} (t) + \frac{i}{\hbar} \left[ \oH_I , \oa_\keps \right]_H (t) .
\end{equation}
Carrying out the remaining commutator with \eqref{eq:HI_adiab} and integrating both equations of motion, we have the coupled integral equations
\begin{align}\label{eq:IntegralEquation1}
    \oa_{\pmu,H} (t) &= \oa_\pmu^{\rm in} e^{-icp(t-t_0)} + C_\pmu \int_{t_0}^t dt' [ \ob_H (t') + \ob_H^\da (t')] e^{-icp(t-t')}, \qquad C_\pmu = \sqrt{\frac{ck}{2\epsilon_0\hbar L^3}} d_0 \dens_{\vk} (\ve_x\cdot\ve_\pmu) e^{-i\vk\cdot\vr_0} ; \\\label{eq:IntegralEquation2}
    \ob_{H} (t) &= \ob^{\rm in} e^{-i\omega_0(t-t_0)} + \int_{t_0}^t dt' \, e^{-i\omega_0(t-t')} 
    \sum_\keps \left[C_\keps \oa_{\keps,H}^\da (t') - C_\keps^* \oa_{\keps,H} (t') \right].
\end{align}
For clarity, we are now denoting the bare (Schr\"{o}dinger-picture) mode operators acting on the separate Hilbert spaces of dipole and field by $\ob^{\rm in}$ and $\oa^{\rm in}_\keps$, as they appear as the initial conditions at $t=t_0$ here.
Next we insert \eqref{eq:IntegralEquation2} into \eqref{eq:IntegralEquation1} to obtain an implicit integral equation for the field mode operators,
\begin{align}\label{eq:DoubleIntegralEq}
    \oa_{\pmu,H} (t) &= \oa_\pmu^{\rm in} e^{-icp(t-t_0)} + C_\pmu \int_{t_0}^t dt' \left[ \ob^{\rm in} e^{-i\omega_0(t'-t_0)} + \ob^{\rm in \da} e^{i\omega_0(t'-t_0)}\right]e^{-icp(t-t')} \nonumber \\
    &+ C_\pmu \int_{t_0}^t dt' e^{-icp(t-t')} \int_{t_0}^{t'} dt'' \left(e^{-i\omega_0(t'-t'')} - c.c.\right) 
    \sum_\keps \left[ C_\keps \oa_{\keps,H}^\da (t'') - h.c. \right] .
\end{align}
Under the usual assumption of weak coupling between scatterer and field modes, we may truncate \eqref{eq:DoubleIntegralEq} at second order in $C_\keps$ and replace the $\oa_{\keps,H} (t'')$ under the double integral by the bare terms $\oa_{\keps}^{\rm in}e^{-ick(t''-t_0)}$. This results in the expansion
\begin{align}\label{eq:DoubleIntegralEq2}
    \oa_{\pmu,H} (t) &\approx \oa_\pmu^{\rm in} e^{-icp(t-t_0)} + \oa_{\pmu,H}^{(1)} (t) + \oa_{\pmu,H}^{(2)} (t),
\end{align}
with the first- and second-order contributions
\begin{align}\label{eq:apmu_1st_order}
    \oa_{\pmu,H}^{(1)} (t) &= C_\pmu \int_{t_0}^t dt' \left[ \ob^{\rm in} e^{-i\omega_0(t'-t_0)} + \ob^{\rm in \da} e^{i\omega_0(t'-t_0)}\right]e^{-icp(t-t')}, \\ \label{eq:apmu_2nd_order}
    \oa_{\pmu,H}^{(2)} (t) &= C_\pmu \int_{t_0}^t dt' \left[  \int_{t_0}^{t'} dt'' \left(e^{-i\omega_0(t'-t'')} - c.c.\right)
    \sum_\keps [C_\keps \oa_\keps^{\rm in \da} e^{ick(t''-t_0)} - h.c.]
    \right]e^{-icp(t-t')}.
\end{align}

\subparagraph{Coherent Amplitudes}\label{sec:CoherentAmplitudes}

To obtain the coherent amplitudes, we take the expectation value of \eqref{eq:DoubleIntegralEq2} with respect to $|\psi^{\rm in}\ra$, $\alpha_\pmu (t) = \la \psi^{\rm in}|\oa_{\pmu,H} (t)|\psi^{\rm in}\ra$. Since the scatterer is initially in the ground state, the $\ob^{\rm in}$ terms vanish: $\la \psi^{\rm in} | \ob^{\rm in} | \psi^{\rm in} \ra = 0$. We also recall that the input pulse amplitudes $\alpha_\keps^{\rm in}$ are defined with respect to the scattering time $t=0$, $\la \psi^{\rm in} | \oa_\keps^{\rm in}|\psi^{\rm in}\ra = \alpha_\keps^{\rm in} e^{-ickt_0}$. 
Hence, we have $\alpha_\pmu(t) \approx \alpha_\pmu^{\rm in} e^{-i c p t} + \alpha_\pmu^{(2)}(t)$, with
\begin{align}\label{eq:2ndOrderAmpl_eq1}
    \alpha_\pmu^{(2)}(t) = C_\pmu \int_{t_0}^t dt'  e^{-icp(t-t')} \int_{t_0}^{t'} dt'' \left(e^{-i\omega_0(t'-t'')} - c.c.\right)
    \sum_\keps \left[ C_\keps \alpha_\keps^{\rm in *} e^{ick t''} - h.c. \right].
\end{align}
Note that the $\keps$-sum is simply the expectation value of the field quadrature at position $\vr_0$ and time $t''$,
\begin{equation}\label{eq:field_at_r0}
    \sum_\keps (C_\keps \alpha_\keps^{\rm in \da} e^{ickt''} - c.c.) = -i d_0 \la \psi^{\rm in}(t'') | \hatvPi(\vr_0) | \psi^{\rm in}(t'') \ra.
\end{equation}
Here, $| \psi^{\rm in}(t'') \ra = e^{-i\oH_0(t''-t_0)/\hbar}| \psi^{\rm in} \ra$ describes the incident light pulse of temporal width $\tau$ propagated from the initial $t_0$ to the time $t''$. Since the center of this pulse is chosen to hit the scatterer position at $t''=0$, the field expectation value \eqref{eq:field_at_r0} vanishes for $|t''| \gg \tau$. Letting 
\begin{equation}\label{eq:h_definition}
    h(t'') \equiv -i d_0 \left(e^{-i\omega_0(t'-t'')} - c.c.\right) \la \psi^{\rm in}(t'') | \hatvPi(\vr_0) | \psi^{\rm in}(t'') \ra
\end{equation}
be the $t''$-integrand function in \eqref{eq:2ndOrderAmpl_eq1}, 
it is then clear that its integral $\int_{t_0}^{t'} dt'' h(t'')$ converges to a finite value $\zeta$ in the limit $t_0 \to -\infty$. In particular, this convergence is uniform over $t' \in (-\infty, t)$, and we claim that
\begin{equation}\label{eq:etaLimitClaim}
    \zeta = \lim_{\eta\to 0} \lim_{t_0\to - \infty} \int_{t_0}^{t'} dt'' e^{\eta t''} h(t''),
\end{equation}
and that convergence in $\eta>0$ is uniform over $t' \in (-\infty, t)$.
To show this, let $\epsilon>0$. By virtue of the triangle and the Cauchy-Schwarz inequalities, 
\begin{align}\label{eq:etaLimitProof}
    \left| \zeta - \int_{-\infty}^{t'} dt'' e^{\eta t''} h(t'') \right|
    &\leq
    \left| \zeta - \int_{T}^{t'} dt'' h(t'') \right| +
    \left| \int_{-\infty}^{T} dt'' e^{\eta t''} h(t'') \right|
    + \left| \int_{T}^{t'} dt'' (1-e^{\eta t''}) h(t'') \right| \nonumber \\
    &\leq
    \left| \zeta - \int_{T}^{t'} dt'' h(t'') \right| +
    e^{\eta T}\left| \int_{-\infty}^{T} dt'' h(t'') \right|
    + |1-e^{\eta T}| \left| \int_{T}^{t'} dt'' h(t'') \right|,
\end{align}
where an arbitrary intermediate time $T < t'$ was introduced. Due to the aforementioned uniform convergence of $\int_{t_0}^{t'} dt'' h(t'')$, there exists a $T_0$ (sufficiently close to $t_0 \to -\infty$, and independent of $t'$) such that both of the first two terms in \eqref{eq:etaLimitProof} are less than $\epsilon/3$ whenever $T \leq \min(T_0, t')$. Having chosen this $T_0$, we set $T=\min(T_0, t')$ then choose a sufficiently small $\eta$ such that the last term is also less than $\epsilon/3$. This choice is independent of $t'$, because either $T_0 < t'$, in which case $T = T_0$ independent of $t'$, or $T_0 \geq t'$, in which case $T=t'$. In the latter case the last term in \eqref{eq:etaLimitProof} is identically zero, so $\eta$ may be chosen freely. In conclusion, this choice of $T$ and $\eta$ is independent of $t'$ and bounds the entire expression by $\epsilon$, proving our claim.

We can now make use of the auxiliary construction \eqref{eq:etaLimitClaim} with $\eta \to 0$ to take the limit $t_0 \to-\infty$ and carry out the integrals in \eqref{eq:2ndOrderAmpl_eq1}. This yields
\begin{equation}\label{eq:alpha_App}
    \alpha_\pmu(t) = \sum_\keps u_{\pmu,\keps} \alpha_\keps^{\rm in}e^{-ickt} + v_{\pmu,\keps} \alpha_\keps^{\rm in *} e^{ickt} + \mathcal O(d_0^4),
\end{equation}
with the transformation coefficients
\begin{align}\label{eq:upk_App}
  u_{\pmu,\keps} &= \frac{\sqrt{kp}}{L^3}
  (\ve_x \cdot \ve_{\keps}) (\ve_x \cdot \ve_{\pmu}) \dens_{\vk} \dens_{\vp}  e^{i(\vk-\vp)\cdot\vr_0}
  \frac{\chi(ck+i0^+)}{p-k-i0^+} + \delta_{\pmu,\keps},
  \\\label{eq:vpk_App}
  v_{\pmu,\keps} &= \frac{\sqrt{kp}}{L^3}
  (\ve_x \cdot \ve_{\keps}) (\ve_x \cdot \ve_{\pmu}) \dens_{\vk} \dens_{\vp}  e^{-i(\vk+\vp)\cdot\vr_0}
  \frac{\chi(ck-i0^+)}{p+k},
\end{align}
and the response function 
\begin{equation}\label{eq:chi}
  \chi(\omega) = \frac{d_0^2}{2\epsilon_0\hbar} \left( \frac{1}{\omega + \omega_0} - \frac{1}{\omega - \omega_0} \right), \qquad \chi (\omega \pm i0^+) = \lim_{\eta \to 0} \chi (\omega \pm i\eta).
\end{equation}
In the off-resonant case $ck\neq\omega_0$, we may omit the $\pm i0^+$ in the argument, resulting in equation \eqreff{eq:uvpk_MainText} in the main text.
The $\delta_{\pmu,\keps}$-term in \eqref{eq:upk_App} represents the zeroth-order contribution of the unscattered field.

For future convenience, let us also calculate the real part of the scatterer's coherent amplitude to leading order, which we obtain by taking the expectation value of \eqref{eq:IntegralEquation2} with respect to $|\psi^{\rm in}\ra$ and using $\ob^{\rm in}|\psi^{\rm in}\ra = 0$:
\begin{align}
    \beta(t) + \beta^*(t) = \la \psi^{\rm in}| \ob_H (t)|\psi^{\rm in} \ra + c.c. = \int_{t_0}^t dt' \, e^{-i\omega_0(t-t')} 
    \sum_\keps \left[C_\keps \alpha_\keps^*(t') - C_\keps^* \alpha_\keps(t') \right]
     + c.c.
\end{align}
To leading order in $d_0$, we replace $\alpha_\keps(t') \approx \alpha_\keps^{\rm in} e^{-ickt'}$, and once again, we can thus identify the field expectation value \eqref{eq:field_at_r0} under the $t'$-integral and leverage \eqref{eq:etaLimitClaim} to introduce the factor $e^{\eta t'}$. We are left with
\begin{align}
    \beta(t) + \beta^*(t) &= \sum_\keps \left(
    \frac{1}{ick+i\omega_0} - \frac{1}{ick-i\omega_0}
    \right)
    C_\keps \alpha_\keps^{\rm in*} e^{ickt} + c.c. = \frac{2\epsilon_0\hbar}{i d_0^2}\sum_\keps [\chi(ck-i0^+)
    C_\keps \alpha_\keps^{\rm in*} e^{ickt} - c.c.]
\end{align}
This yields the relation
\begin{align}\label{eq:beta_relation}
    \frac{i d_0 (\ve_x \cdot \ve_\pmu) \dens_{\vp}}{\sqrt{2\eps_0 \hbar c p L^3}} e^{-i\vp\cdot\vr_0} [\beta(t) + \beta^*(t)] = \frac{e^{-i\vp\cdot\vr_0}}{L^3} (\ve_x \cdot \ve_\pmu) \dens_{\vp} \sum_\keps (\ve_x \cdot \ve_\keps) \dens_{\vk} \left[
    \chi(ck-i0^+)\alpha_\keps^{\rm in*} e^{ickt} - c.c.
    \right],
\end{align}
which will be useful when computing the coherent amplitudes in different gauge representations.

\subparagraph{Covariance Matrix}\label{sec:CovarianceMatrix}

With the time-evolved Heisenberg-picture mode operators at hand, we can not only evaluate the mean coherent amplitudes, but also the second moments, i.e., covariances. This is all we need here since the system remains Gaussian at all times due to the Gaussian initial state and the quadratic Hamiltonian. .

The covariance matrix, contains all second-order cumulants between all combinations of the mode operators and their hermitean conjugates, reflects the vacuum properties of the state and does not depend on any of the coherent displacements. In the absence of coupling between the modes, it would simply reduce to the identity matrix. Hence, we can expand it perturbatively around the identity in the weak-coupling regime considered here. 

Let us, for the moment, introduce the shorthand notation $\oa_n$ with $n=\keps,b$ subsuming any of the mode operators $\oa_\keps$ or $\oa_b \equiv \ob$. Given a (Gaussian) quantum state $\varrho(t)$ with mean displacements $\alpha_n (t) = \tr [\varrho (t) \oa_n] $, the covariance matrix can be expressed as~\cite{Safranek2017} 
\begin{equation}\label{sigmaDefinition}
  \sigma = \begin{pmatrix}
  \Xi&\Upsilon\\\Upsilon^*&\Xi^*
  \end{pmatrix},
\end{equation}
with 
\begin{align}
    \Xi_{nm} &= \tr \left[ \varrho(t) \{ \oa_n - \alpha_n (t), \oa_m^\da - \alpha_m^* (t) \}  \right] = 2 \la \psi_{\rm in} | [\oa_{m,H}^\da (t) - \alpha_m^* (t)][\oa_{n,H}(t) - \alpha_n (t)]|\psi_{\rm in} \ra + \delta_{nm} , \nonumber \\
    \Upsilon_{nm} &= \tr \left[ \varrho(t) \{ \oa_n - \alpha_n (t), \oa_m - \alpha_m (t) \}  \right] = 2 \la \psi_{\rm in} | [\oa_{n,H}(t) - \alpha_n (t)][\oa_{m,H} (t) - \alpha_m (t)]|\psi_{\rm in} \ra . \label{eq:Covariances_eq2}
\end{align}
Here, we only access the submatrices of the field degrees of freedom, $(n,m)=(\keps,\pmu)$. Inserting the perturbative weak-coupling expansion \eqref{eq:DoubleIntegralEq2} of the mode operators, we arrive at 
\begin{align}
    \Xi_{\keps, \pmu} =& \, \delta_{\keps,\pmu} + 2 \la \psi_{\rm in} | \oa_{\keps,H}^{(1)\da}(t) \oa_{\pmu,H}^{(1)}(t) | \psi_{\rm in} \ra \label{eq:Xi_weakCoupling} \\
    &+ 2 \left\la \psi_{\rm in} \left| \left[\oa_{\keps,H}^{(2)\da}(t) - \alpha_\keps^{(2)*}(t)\right] (\oa_\pmu^{\rm in}e^{icpt_0} - \alpha_\pmu^{\rm in})e^{-icpt} + (\oa_\keps^{\rm in \da}e^{-ickt_0} - \alpha_\keps^{\rm in *})e^{ickt}\left[\oa_{\pmu,H}^{(2)}(t) - \alpha_\pmu^{(2)}(t)\right] \right| \psi_{\rm in} \right\ra, \nonumber \\
    \Upsilon_{\keps, \pmu} =& \, 2 \la \psi_{\rm in} | \oa_{\keps,H}^{(1)}(t) \oa_{\pmu,H}^{(1)}(t) | \psi_{\rm in} \ra \label{eq:Ups_weakCoupling} \\
    &+ 2 \left\la \psi_{\rm in} \left| \left[\oa_{\keps,H}^{(2)}(t) - \alpha_\keps^{(2)}(t)\right] (\oa_\pmu^{\rm in}e^{icpt_0} - \alpha_\pmu^{\rm in})e^{-icpt} + (\oa_\keps^{\rm in}e^{ickt_0} - \alpha_\keps^{\rm in})e^{-ickt} \left[\oa_{\keps,H}^{(2)}(t) - \alpha_\keps^{(2)}(t)\right] \right| \psi_{\rm in} \right\ra, \nonumber 
\end{align}
which are both of second order in the weak coupling, i.e., valid up to $\mathcal O(d_0^4)$.
Here, we have exploited that $\alpha_\pmu^{(1)}(t) = \la \psi_{\rm in} | \oa_{\pmu,H}^{(1)} | \psi_{\rm in} \ra = 0$, because $\la \psi_{\rm in} | \ob^{\rm in} | \psi_{\rm in} \ra = 0$. 
Moreover, since $(\oa_\keps^{\rm in}e^{ickt_0} - \alpha_\keps^{\rm in})|\psi^{\rm in}\ra = 0$, it follows that the entire second line in \eqref{eq:Xi_weakCoupling} vanishes, as well as the first half of the second line in \eqref{eq:Ups_weakCoupling}. Substituting \eqref{eq:apmu_1st_order} and \eqref{eq:apmu_2nd_order}, commuting the mode operators, and performing the remaining time integrals yields the explicit matrix elements
\begin{align}
    \Xi_{\keps,\pmu} - \delta_{\keps,\pmu}&= 2 C_\keps^* C_\pmu \frac{e^{-i\omega_0(t-t_0)} - e^{ick(t-t_0)}}{-ick-i\omega_0} \frac{e^{i\omega_0(t-t_0)} - e^{-icp(t-t_0)}}{icp+i\omega_0} \\
    \Upsilon_{\keps,\pmu} &= 2 C_\keps C_\pmu \frac{e^{-i\omega_0(t-t_0)} - e^{-ick(t-t_0)}}{ick-i\omega_0} \frac{e^{i\omega_0(t-t_0)} - e^{-icp(t-t_0)}}{icp+i\omega_0}, \\
    &\quad + 2 C_\keps C_\pmu \left[
    \frac{1}{ick+i\omega_0} \left( \frac{1 - e^{-ic(p+k)(t-t_0)}}{icp+ick} - \frac{e^{-i\omega_0(t-t_0)-ick(t-t_0)} - e^{-ic(p+k)(t-t_0)}}{icp-i\omega_0} \right) \right. \nonumber \\
    &\qquad \qquad \qquad \left. - \frac{1}{ick-i\omega_0} \left( \frac{1 - e^{-ic(p+k)(t-t_0)}}{icp+ick} - \frac{e^{i\omega_0(t-t_0)-ick(t-t_0)} - e^{-ic(p+k)(t-t_0)}}{icp+i\omega_0} \right)
    \right] \nonumber 
\end{align}
In order to take the limit $t_0 \to -\infty$, note that in the end, the covariance matrices will be applied to coherent amplitude vectors representing pulses with a finite temporal width. In our case, we will have terms such as $\sum_{\keps,\pmu} [\partial \alpha_\keps^* (t)/\partial\theta_j] \Xi_{\keps,\pmu} [\partial \alpha_\pmu (t)/\partial \theta_l]$, evaluated in the continuum limit and at finite $t$; see Supplementary Section~\ref{sec:QFI_Appendix} below. Any contribution that oscillates with $e^{\pm ickt_0}$ or $e^{\pm icpt_0}$ will thus converge to zero as $t_0 \to -\infty$. The residual time-independent covariances representing the squeezed mode vacuum of the weakly coupled scatterer and light field are
\begin{align} \label{eq:XiMatrixElem_eq1}
    \Xi_{\keps, \pmu} - \delta_{\keps,\pmu} &= 2 \frac{C_\pmu}{cp + \omega_0} \frac{C_\keps^*}{ck + \omega_0}, 
    \\
    \Upsilon_{\keps, \pmu} &= 2 \frac{C_\keps}{\omega_0 - ck} \frac{C_\pmu}{cp + \omega_0}
    - 2 \frac{C_\keps C_\pmu}{cp + ck} \left[
    \frac{1}{ ck + \omega_0} - \frac{1}{ck-\omega_0}
    \right] = - 2\frac{C_\pmu C_\keps}{ck+cp} \left(
    \frac{1}{cp+\omega_0} + \frac{1}{ck+\omega_0}
    \right) .
    \label{eq:UpsMatrixElem_eq1}
\end{align}

\clearpage 

\section{Quantum Fisher Information}
\label{sec:QFI_Appendix}

Here we provide details on the calculation leading to the QFI matrix of the field state at a given time $t$ with respect to the scatterer parameters $\vtheta$. As the state is Gaussian, the QFI can be expressed in terms of mean displacements, covariances, and derivatives thereof with respect to the parameters. We will give the relevant expressions in the continuum limit, which we have used in our numerical evaluation of the QFI. 

\subparagraph{Gaussian State QFI}

The quantum Fisher information matrix $\mathcal J$ of a multimode Gaussian state with respect to some parameters $\vtheta$ [Eq.~(\reff{GaussianQFI}) in the main text] depends on both the vector of all coherent displacements $\valpha (\boldsymbol\theta)$ and on the covariance matrix $\sigma (\boldsymbol\theta)$ defined in (\ref{sigmaDefinition}); see Ref.~\cite{Safranek2017}, which also contains the explicit form of the here omitted vacuum contribution $\mathcal V$. The latter depends neither on the displacements nor on time, and it has the same value regardless of whether any coherent light scattering occurs at all.

The dominant contribution comes from the parameter sensitivity of the displacements $\valpha (\vtheta)$, compactly written as the bilinear form \eqreff{GaussianQFI} in the main text,  reflecting that this information about the scatterer is obtainable by measuring those coherent amplitudes and subsequently deducing an estimate for $\boldsymbol\theta$. The expression also depends on the inverse of the covariance matrix $\sigma$, which we can approximate in the weak-coupling regime by expanding it around identity,
\begin{equation}
    \pmat{ \Xi & \Upsilon \\ \Upsilon^* & \Xi^* }^{-1} \approx \id - \pmat{ \Xi - \id & \Upsilon \\ \Upsilon^* & \Xi^* - \id },
\end{equation} 
which leads to the second line of \eqreff{GaussianQFI} in the main text.
Indeed, at vanishing scatterer-field coupling, we have $\Xi \to \id$ and $\Upsilon \to 0$, and the leading-order expansion spares us the effort of performing a numerical matrix inversion.

\subparagraph{Derivatives of the Amplitude}

To compute the QFI, we need to take derivatives of the coherent amplitudes (\ref{eq:alpha_App}) with respect to the parameters of interest. This is tedious but not difficult; we will now state the essential steps in the continuum limit $L\to\infty$. The derivatives with respect to the scatterer's coordinates $\theta_j = \vr_0 \cdot\ve_j$ and the off-resonant polarizability $\theta_0 = \chi_0 = d_0^2/\hbar \epsilon_0 \omega_0$ are
\begin{align}\label{eq:dAlphaDR_eq1}
  \frac{\partial  \alpha_\pmu(t)}{\partial \theta_{j\neq0}}\Big|_{\boldsymbol r_0=0} &= \sum_\keps \left[ \frac{\partial u_{\pmu,\keps}^*}{\partial \theta_j} \alpha^{\rm in}_\keps + \frac{\partial v_{\pmu,\keps}}{\partial \theta_j} \alpha_\keps^{\rm in*} \right]
  \\
  &\to i\sqrt{p} (\ve_\pmu \cdot \ve_x) \dens_{\vp} \int_0^\infty \frac{dk}{2\pi} \sqrt{k} \dens_{\vk} 
  \left[
  \frac{\chi( ck+i0^+)\alpha^{\rm in}(k, t)}{ k- p+i0^+} (p_j - k\delta_{jz}) +
  \frac{\chi(ck-i0^+)\alpha^{\rm in *}(k, t)}{ k+ p} (p_j + k\delta_{jz})
  \right] 
  \nonumber\\\nonumber\\ \label{eq:dAlphaDChi_eq1}
  \frac{\partial \alpha_\pmu(t)}{\partial \theta_0}\Big|_{\boldsymbol r_0=0}  &= \sum_\keps \left[ \frac{\partial u_{\pmu,\keps}^*}{\partial \theta_0} \alpha^{\rm in}_\keps + \frac{\partial v_{\pmu,\keps}}{\partial \theta_0} \alpha_\keps^{\rm in *} \right]
  \\
  &\to - \sqrt{p} (\ve_\pmu \cdot \ve_x) \dens_{\vp}  \int_0^\infty \frac{dk}{2\pi} \sqrt{k} \dens_{\vk}  \left[
  \frac{\partial\chi( ck+i0^+)}{\partial \theta_0}\frac{\alpha^{\rm in}(k, t)}{ k- p+i0^+}
  +
  \frac{\partial\chi( ck-i0^+)}{\partial \theta_0}\frac{\alpha^{\rm in *}(k, t)}{ k+ p}
  \right]
  \nonumber
\end{align}
Here, the derivatives are evaluated at the current reference position of the scatterer,  $\vr_0=0$. Also, we use that the incident light propagates along the $z$-direction and is $x$-polarized, $\alpha_\keps^{\rm in} = \alpha_{k\ve_z}^{\rm in} \delta_{k_x 0} \delta_{k_y 0} \delta_{\eps 1}$  with $\ve_{\vk 1} = \ve_x$ and $k>0$. In the continuum limit, this translates to $ \sum_\keps \alpha_\keps^{\rm in} \to \sum_{\eps} (L/2\pi) \int dk \, \alpha_{k\ve_z}^{\rm in} \delta_{\eps 1}$, and we define the amplitude density per unit area, $ \alpha^{\rm in} (k,t) = (\alpha_{k\ve_z}^{\rm in}/L^2) e^{-ickt}$, which is nonzero strictly only for $k>0$.
If we use the parameterization
\begin{equation}\label{kepsParameterization}
    \ve_{\vp 1} = \frac{\vp\times \ve_x}{|\vp\times\ve_x|}, \qquad 
    \ve_{\vp 2} = \frac{\vp}{p} \times \ve_{\vp 1}, \qquad  
    \vp = p(\cos\vartheta_p \ve_z + \cos\varphi_p\sin\vartheta_p\ve_x + \sin\varphi_p\sin\vartheta_p\ve_y),
\end{equation}
the last factor turns into $\ve_\pmu \cdot \ve_x = -\delta_{\mu 2} \sqrt{\sin^2\vartheta_p\sin^2\varphi_p+\cos^2\vartheta_p}$, so that
\begin{align}\label{eq:dAlphadR_eq2}
  \frac{\partial \alpha_\pmu(t)}{\partial \theta_{j\neq 0}}  &= \frac{1}{i} \left[ \delta_{jz} f_1(p,t) + \frac{p_j}{p}\, f_2(p,t)
  \right] \delta_{\mu 2} \sqrt{\sin^2\vartheta_p\sin^2\varphi_p+\cos^2\vartheta_p}, \\
\label{eq:dAlphaDChi_eq2}
  \frac{\partial \alpha_\pmu(t)}{\partial \theta_0}  &= f_3(p,t)
  \delta_{\mu 2} \sqrt{\sin^2\vartheta_p\sin^2\varphi_p+\cos^2\vartheta_p}.
\end{align}
Herein, the $f$'s abbreviate the frequency integrals 
\begin{equation}\begin{aligned}\label{eq:f123}
    f_1(p,t) &= p^{1/2} \dens_{\vp} \int_0^\infty \frac{dk}{2\pi} k^{3/2} \dens_{\vk} \left[
    \frac{\chi( ck-i0^+) \alpha^{\rm in *}(k,t)}{k+p} -
    \frac{\chi( ck+i0^+)\alpha^{\rm in}(k,t)}{k-p+i0^+}
    \right],
    \\\\
    f_2(p,t) &=
    p^{3/2} \dens_{\vp} \int_0^\infty \frac{dk}{2\pi} k^{1/2}\dens_{\vk} \left[
    \frac{\chi( ck-i0^+) \alpha^{\rm in *}(k,t)}{k+p} +
    \frac{\chi( ck+i0^+) \alpha^{\rm in}(k,t)}{k-p+i0^+}
    \right],
    \\\\
    f_3(p,t) &= p^{1/2} \dens_{\vp} \int_0^\infty \frac{dk}{2\pi} k^{1/2} \dens_{\vk} \left[
    \frac{\partial\chi(ck-i0^+)}{\partial \chi_0} \frac{\alpha^{\rm in *}(k,t)}{k+p} +
    \frac{\partial\chi(ck+i0^+)}{\partial \chi_0} \frac{\alpha^{\rm in}(k,t)}{k-p+i0^+}
    \right].
\end{aligned}\end{equation}
Lastly, from (\ref{eq:chi}), we immediately obtain the remaining derivative of the response function,
\begin{equation}\label{eq:chi0_derivative_weakCoupling}
  \frac{\partial\chi(\nu)}{\partial \theta_0} = \frac{\partial\chi(\nu)}{\partial \chi_0} = \frac{\chi(\nu)}{\chi_0} \approx 1.
\end{equation}
The last step amounts to the off-resonance approximation $\chi(\nu) \approx \chi_0$. It implies that $\partial \alpha_\pmu (t)/\partial \theta_0 \approx [\alpha_\pmu (t)-\alpha_\pmu^{\rm in}]/\chi_0$.
This is as far as we are able to go analytically. The frequency integrals $f_{1,2,3}$ must be evaluated numerically.

We can now proceed to calculate the QFI, omitting the vacuum contribution $\mathcal V$. 
We begin by explicitly expanding the bilinear form  \eqreff{GaussianQFI} in the main text:
\begin{equation}
\begin{aligned}\label{GaussianQFISummands}
    2\frac{\partial \valpha^*}{\partial \theta_{j}} \cdot \Xi \frac{\partial \valpha}{\partial \theta_{l}} + c.c. &= 4\text{Re} \sum_{\pmu,\qmu} \frac{\partial \alpha_\pmu^*}{\partial\theta_j} \frac{\partial \alpha_\qmu}{\partial\theta_l} \Xi_{\pmu,\qmu}, 
    \\
    2\frac{\partial \valpha^*}{\partial \theta_{j}} \cdot \Upsilon \frac{\partial \valpha}{\partial \theta_{l}} + c.c. &= 4\text{Re} \sum_{\pmu,\qmu} \frac{\partial \alpha_\pmu^*}{\partial\theta_j} \frac{\partial \alpha_\qmu}{\partial\theta_l} \Upsilon_{\pmu,\qmu}
\end{aligned}
\end{equation}
In the following, we calculate the above expressions for all combinations of indices $j, l$.

\subparagraph{Position Estimation}

\begin{figure}[!htb]
\begin{overpic}[width=\linewidth]{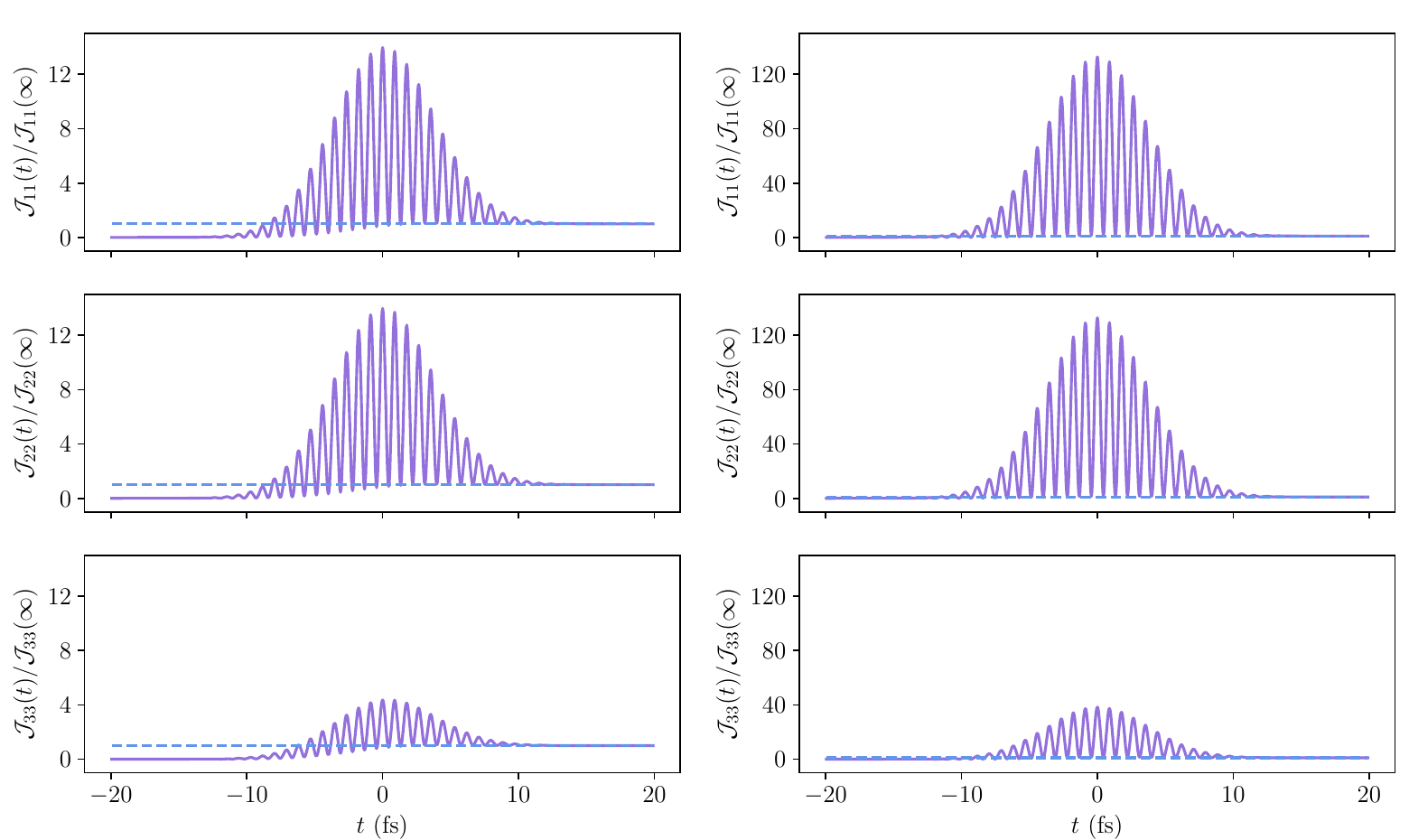}
  \put (8, 16) {\large (c)}
  \put (8, 34) {\large (b)}
  \put (8, 53) {\large (a)}
  \put (59, 16) {\large (f)}
  \put (59, 34) {\large (e)}
  \put (59, 53) {\large (d)}
\end{overpic}
\caption{
QFI for estimating the components of $\vr_0$, for a scatterer with radius $a_0 = \lambdain/30$ (a-c) and $a_0 = \lambdain/53$ (d-f). The QFI is normalized to the  far-field value $\mathcal J_{\ell\ell}(\infty)$ and the vacuum contribution $\mathcal V$ is omitted. The horizontal axis shows time in $\si{fs}$. All data were obtained with a polarizability $\chi_0 = \SI{13.0}{nm^3}$ at $\lambdain = \SI{532}{nm}$, i.e. the smaller scatterer has a different polarizability density. Panel (a) is identical to Fig.~\reff{fig:QFI}(a) in the main text. The far-field values are independent of $a_0$ as long as $a_0 \ll \lambdain$, i.e., the same for both scatterer sizes. Different particle sizes cannot be distinguished in the far field if the polarizabilities are the same.
}
\label{fig:QFI_r0_all}
\end{figure}

For $j, l \neq 0$, the estimation parameters are the position coordinates of the scatterer, $\theta_j = \vr_0\cdot\ve_j$ and $\theta_l = \vr_0\cdot\ve_l$. Substitution of \eqref{eq:dAlphadR_eq2}, \eqref{eq:XiMatrixElem_eq1} and \eqref{eq:UpsMatrixElem_eq1} into \eqref{GaussianQFISummands} gives
\begin{align}
    2\frac{\partial \valpha^*}{\partial \theta_{j}} \cdot \Xi \frac{\partial \valpha}{\partial \theta_{l}} + c.c. \xrightarrow[L\to\infty]{} & \, 4\text{Re} \bigg[
    \int \frac{d\Omega_p}{(2\pi)^2} (\sin^2\vartheta_p\sin^2\varphi_p + \cos^2\vartheta_p)\nonumber\\
    &\times\int \frac{dp}{2\pi} p^2
    \left\{
    \delta_{j3} f_1(p,t) + \frac{p_j}{p}\, f_2(p,t)
    \right\}^*\left\{
    \delta_{l3} f_1(p,t) + \frac{p_l}{p}\, f_2(p,t)
    \right\}\nonumber\\
    &\quad+ \int \frac{d\Omega_p d\Omega_{p'}}{(2\pi)^4} (\sin^2\vartheta_{p'}\sin^2\varphi_{p'} + \cos^2\vartheta_{p'}) (\sin^2\vartheta_p\sin^2\varphi_p + \cos^2\vartheta_p)
    \nonumber\\
    &\times\int \frac{dpdp'}{(2\pi)^2} p'^2 p^2
    \left\{
    \delta_{j3} f_1(p',t) + \frac{p_j'}{p}\, f_2(p',t)
    \right\}^* \left\{
    \delta_{l3} f_1(p,t) + \frac{p_l}{p}\, f_2(p,t)
    \right\} \delta \Xi(p',p)
    \bigg]
    \\\nonumber\\
    2\frac{\partial \valpha^*}{\partial \theta_{j}} \cdot \Upsilon \frac{\partial \valpha}{\partial \theta_{l}}  + c.c. \xrightarrow[L\to\infty]{} & \, 4\text{Re} \bigg[
    \int \frac{d\Omega_p d\Omega_{p'}}{(2\pi)^4} (\sin^2\vartheta_{p'}\sin^2\varphi_{p'} + \cos^2\vartheta_{p'}) (\sin^2\vartheta_p\sin^2\varphi_p + \cos^2\vartheta_p)
    \nonumber\\
    &\times\int \frac{dpdp'}{(2\pi)^2} p'^2 p^2
    \left\{
    \delta_{j3} f_1(p',t) + \frac{p_j'}{p'}\, f_2(p',t)
    \right\}^* \left\{
    \delta_{l3} f_1(p,t) + \frac{p_l}{p}\, f_2(p,t)
    \right\}^* \Upsilon(p',p)
    \bigg].
\end{align}
Recall that the $p_j$ components in the curly brackets depend on the integration angle, c.f.~\eqref{kepsParameterization}. The integrals simplify drastically, because any odd $p_j$-term will integrate to zero. In the first expression, only the summands with $\delta_{j3}\delta_{l3}$ or $p_j^2$ under the integral survive. In the $\delta\Xi$ and $\Upsilon$ expressions, all terms vanish except for the one with $\delta_{j3}\delta_{l3}$.
The angular integrals over the remaining terms can be done analytically, leaving only the frequency integrals. Putting everything together, we have
\begin{equation}\begin{aligned}\label{eq:QFI_final_r0}
    \mathcal J_{11}(t) - \mathcal V_{11} &= \frac{8}{15\pi} \int_0^\infty \frac{dp}{2\pi}\ p^2 \left|
    f_2(p,t)
    \right|^2 = \frac{\mathcal J_{22}(t) - \mathcal V_{22}}{2},
    \\
    \mathcal J_{33}(t) - \mathcal V_{33} &= 2(\mathcal J_{11}(t) - \mathcal V_{11}) + \frac{8}{3\pi} \int_0^\infty \frac{dp}{2\pi} p^2 \left| f_1(p,t)\right|^2 \\
    &+ \frac{8}{9\pi^2}
      \int_0^\infty \frac{dp'dp}{(2\pi)^2} p'^2 p^2 \left(
      f_1^*(p',t)f_1^*(p,t)
      \Upsilon(p',p)
      -
      f_1^*(p',t)f_1(p,t)
      \delta \Xi(p',p) + c.c.
      \right),
\end{aligned}\end{equation}
and $\mathcal J_{jl} - \mathcal V_{jl} = 0$ for $j \neq l$. The three diagonal entries, which represent the Cram\'{e}r-Rao precision bounds for estimating $x_0,y_0,z_0$, are plotted in Fig.~\ref{fig:QFI_r0_all} for two scatterer sizes. Panel (e) corresponds to Fig.~\reff{fig:QFI}(a) in the main text.

\subparagraph{Polarizability Estimation}

For $j,l=0$, the estimation parameter is the weak-coupling polarizability, $\theta_0 = \chi_0$. Substituting \eqref{eq:dAlphaDChi_eq2}, \eqref{eq:XiMatrixElem_eq1} and \eqref{eq:UpsMatrixElem_eq1} into \eqref{GaussianQFISummands}, and performing a calculation similar to the one above, we obtain the diagonal entry of the QFI matrix corresponding to polarizability estimation:
\begin{equation}\label{eq:QFI_final_chi}
    \mathcal J_{00} (t) - \mathcal V_{00}
    = 4 \int_0^\infty \frac{dp}{3\pi^2} p^2 \left\{ \left| f_3(p,t)\right|^2 - 
    \int_0^\infty \frac{dp'}{6\pi^2} p'^2 \left[f_3^*(p',t)f_3^*(p,t)
    \Upsilon(p',p)
    +
    f_3^*(p',t)f_3(p,t)
    \delta \Xi(p',p) + c.c. \right] \right\} .
\end{equation}
This is the quantity shown in Fig.~\reff{fig:QFI}(b) in the main text.

At weak, off-resonant coupling, the covariance matrix terms $\delta\Xi$ and $\Upsilon$ will give only minor contributions to the QFI, as confirmed by our numerical assessment. Neglecting them, we can approximate
\begin{equation}
    \mathcal J_{00} (t) - \mathcal V_{00} \approx 4 \sum_\pmu \left|\frac{\partial \alpha_\pmu (t)}{\partial \theta_0} \right|^2 \approx \frac{4}{|\chi_0|^2} \sum_\pmu \left| \alpha_\pmu(t) - \alpha_\pmu^{\rm in} \right|^2 \equiv \frac{4 N^{\rm sc} (t)}{|\chi_0|^2},
\end{equation}
where $N^{\rm sc} (t)$ is the number of photons in the scattered field at a given time $t$.
In other words, the QCRB for polarizability estimation, $\Delta \chi_0 /\chi_0 = 1/2\sqrt{N^{\rm sc}}$ from Eq.~\eqreff{FarFieldQFI} in the main text, is not only valid in the far field, but for any point in time during the scattering process.

\subparagraph{Position--Polarizability Covariance}

The only remaining entries are off-diagonal ones with $j=0, l>0$. Substituting \eqref{eq:dAlphaDChi_eq2}, \eqref{eq:dAlphadR_eq2}, \eqref{eq:UpsMatrixElem_eq1} and \eqref{eq:XiMatrixElem_eq1} in \eqref{GaussianQFISummands} gives vanishing $\mathcal J_{01}(t) - \mathcal V_{01} = 0$ and $\mathcal J_{02}(t) - \mathcal V_{02} = 0$, while
\begin{equation}\label{QFI_final_offdiag}\begin{aligned}
    \mathcal J_{03}(t) - \mathcal V_{03}
    &= 2i \int_0^\infty \frac{dp}{3\pi^2} p^2 \left \{ f_1(p,t)f_3^*(p,t)
    + \int_0^\infty \frac{dp'}{3\pi^2} p'^2 \left[ f_1^*(p',t)f_3^*(p,t)
    \Upsilon(p',p)
    +
    f_1^*(p',t)f_3(p,t)
    \delta \Xi(p',p)\right] \right\} + c.c.
\end{aligned}\end{equation}

\subparagraph{Numerical Methods}

The integrals in \eqref{eq:QFI_final_r0}, \eqref{eq:QFI_final_chi}, and \eqref{QFI_final_offdiag} were computed by discretizing wave number (frequency) space based on $p_n = d \sinh(\Delta[n - n_0]) + k_0$, 
with $n$ an integer ranging from 0 to $\mathcal N$, and 
$n_0$ chosen such that $p_{-1} < 0 \leq p_0$. This parameterization ensures that the resolution becomes coarser as one moves away from $k_0$, the wave number the incident wave packet is centered on. The scaling parameters were chosen as $d/q = \SI{2.5e-3}{}$ and $\Delta = \SI{3.8e-2}{}$. The maximal index $\mathcal N$ was chosen so $p_{\mathcal N} = \SI{1.1e+3}{}q \gg 2\pi/a_0$, ensuring that the hard cutoff imposed by the constraint $n < \mathcal N$ has no effect. The functions $f_i(p)$ defined in \eqref{eq:f123} were calculated using the Sokhotski--Plemelj theorem,
\begin{equation}
    \frac{1}{\nu + i0+} = \mathcal P \frac{1}{\nu} - i\delta(\nu).
\end{equation}
$\mathcal P$ denotes the Cauchy principal value. The $\delta(\nu)$ term was evaluated analytically in \eqref{eq:f123}, while the principal value integral was computed using the QUADPACK routine~\cite{Piessens1983}.

\begin{figure}[!htb]
\includegraphics[width=\linewidth]{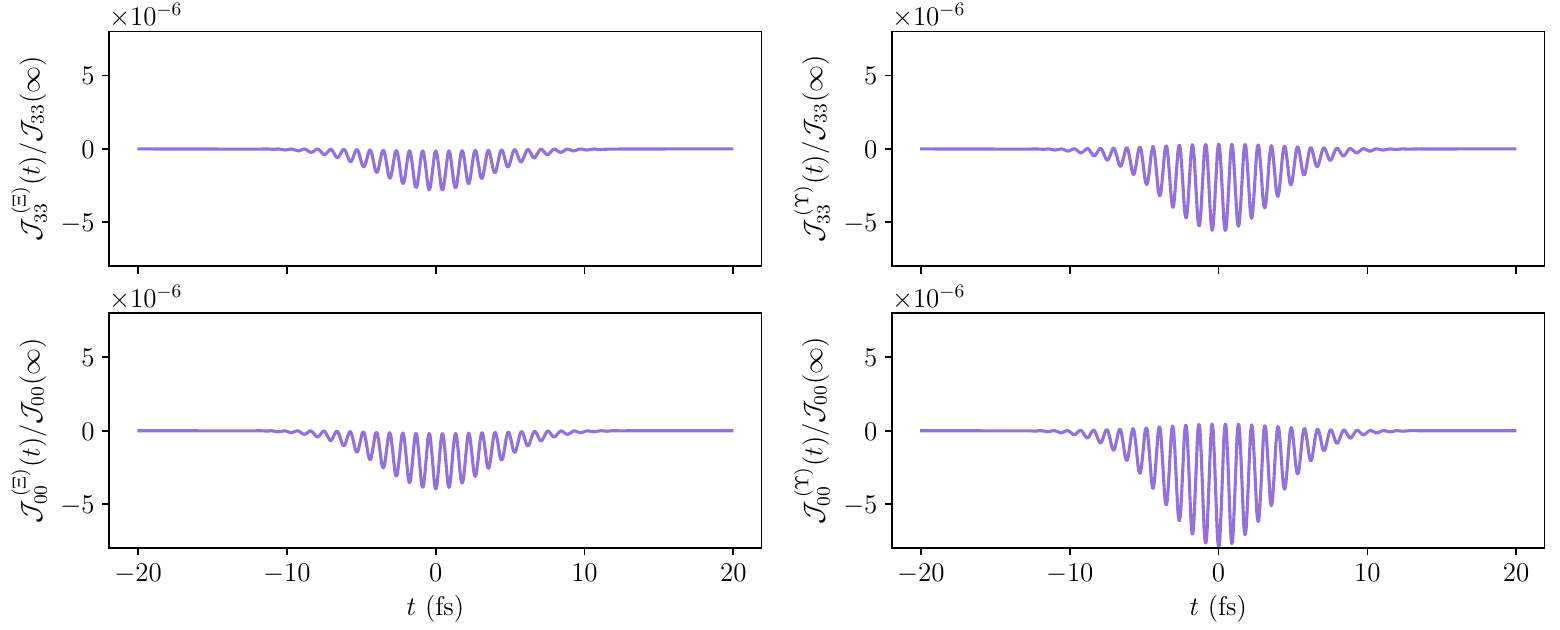}
\caption{
If the covariance matrix $\sigma$ defined in \eqref{sigmaDefinition} is close to, but different from $\id$, the QFI is modified by the corrections $\mathcal J^{(\Xi)} = \partial \valpha^* \cdot \delta\Xi \partial \valpha$ and $\mathcal J^{(\Upsilon)} = \partial \valpha^* \cdot \Upsilon \partial \valpha$.
The data shown are for $\chi_0 = \SI{13.0}{nm^3}$, $\lambdain=\SI{532}{nm}$, and $a_0=\lambdain/30$. The $\mathcal O(10^{-6})$ magnitude of the correction confirms that effects from the dressed electromagnetic vacuum are negligible at this coupling strength.
}
\label{fig:QFI_corrections}
\end{figure}

Our numerical evaluation could show that, in the off-resonant weak-coupling regime we consider here, the covariance matrix terms $\delta \Xi, \Upsilon \neq 0$ describing the squeezing of the mode vacuum due to the presence of the scatterer give rise to merely negligible corrections to the QFI. Curiously, these corrections only appear in $\mathcal J_{03}$, $\mathcal J_{00}$ and $\mathcal J_{33}$; we plot them for our parameter settings in Fig.~\ref{fig:QFI_corrections}. A quick comparison to the values in Fig.~\reff{fig:QFI} of the main text and in Fig.~\ref{fig:QFI_r0_all} confirms that the corrections are indeed negligible.

\subparagraph{Asymptotics}

Finally, let us remark on the asymptotic behavior of the QFI at large $k_M = 1/a_0$. The dominant integrals in \eqref{eq:QFI_final_r0}, \eqref{eq:QFI_final_chi} and \eqref{QFI_final_offdiag} have the form 
\begin{equation}
\mathcal J_{jl} - \mathcal V_{jl} = \int_0^\infty dp p^2 |\dens_{\vp}|^2 \mathcal F_{jl}(p),    
\end{equation}
where $\mathcal F_{jl}(p)$ is a smooth function of $p$. If $\mathcal F_{jl}(p) \sim p^n$ for large $p$, then the integral will scale like $1/a_0^{n+3}$ for small $a_0$, as can be observed by performing a change of variables $p \to a_0 p$. Using that $f_1 \sim p^{-1/2}, f_2 \sim p^{1/2}, f_3 \sim p^{-1/2}$, we can conclude $\mathcal J_{00} - \mathcal V_{00} \sim (\lambdain/a_0)^2$ for polarizability estimation, $\mathcal J_{jj} - \mathcal V_{jj} \sim (\lambdain/a_0)^4$ for position estimation ($j=1,2,3$), and $\mathcal J_{03} - \mathcal V_{03} \sim (\lambdain/a_0)^3$ for the non-zero off-diagonal term. Here, $\lambdain$ is the characteristic wavelength of the incident light pulse.

\clearpage

\section{Far Field and Scattering Cross Section}
\label{sec:FarFieldAmplitudes}

Here, we derive the far-field scattering amplitudes \eqreff{FarFieldAmplitude} in the main text, as well as the associated scattering cross section and total number of scattered photons $N^{\rm sc}$. The latter serves as a normalization constant in the quantum and classical Cram\'{e}r-Rao bounds on the estimation precision of the scatterer's parameters. 

We follow along the lines of the full scattering calculation in Supplementary Section~\ref{sec:TimeEvol}. Rather than integrating the Heisenberg equations of motion to a finite time $t$ as in \eqref{eq:IntegralEquation1} and \eqref{eq:IntegralEquation2}, we now integrate to $t \to \infty$. 
Specifically, we let $t \to \infty$ in the scattering contribution to the coherent amplitudes \eqref{eq:2ndOrderAmpl_eq1}:
\begin{align}\label{eq:2ndOrderAmpl_farField}
    \alpha_\pmu^{(2)}(t) = C_\pmu \int_{t_0}^\infty dt'  e^{-icp(t-t')} \int_{t_0}^{t'} dt'' \left(e^{-i\omega_0(t'-t'')} - c.c.\right)
    \sum_\keps \left[ C_\keps \alpha_\keps^{\rm in *} e^{ick t''} - h.c. \right].
\end{align}
Again, let $h(t'')$ be the integrand function as defined in \eqref{eq:h_definition}, which due to \eqref{eq:field_at_r0} describes a finite-time pulse at the scatterer position. Analogously to the auxiliary limit construction \eqref{eq:etaLimitClaim}, we now claim  
\begin{equation}\label{eq:etaLimitClaim2}
    \zeta = \lim_{\eta\to 0} \lim_{t_0\to - \infty} \int_{t_0}^{t'} dt'' e^{-\eta |t''|} h(t''),
\end{equation}
with uniform convergence in $\eta>0$ over $t' \in \mathbb R$. Notice the absolute value $|t''|$ in the exponent, which differs from the previous claim \eqref{eq:etaLimitClaim}. Nevertheless, the proof proceeds similarly to \eqref{eq:etaLimitProof} by means of the Cauchy-Schwartz and the triangle inequality:
\begin{align}\label{eq:etaLimitProof2}
    \left| \zeta - \int_{-\infty}^{t'} dt'' e^{\eta t''} h(t'') \right|
    &\leq
    \left| \zeta - \int_{T}^{t'} dt'' h(t'') \right| +
    \left| \int_{-\infty}^{T} dt'' e^{-\eta |t''|} h(t'') \right|
    + \left| \int_{T}^{t'} dt'' (1-e^{-\eta |t''|}) h(t'') \right| \nonumber \\
    &\leq
    \left| \zeta - \int_{T}^{t'} dt'' h(t'') \right| +
    \left| \int_{-\infty}^{T} dt'' h(t'') \right|
    + |1-e^{-\eta |T|}| \left| \int_{T}^{t'} dt'' h(t'') \right|,
\end{align}
where $T<t'$ is again arbitrary. In the same manner as before, $T$ and $\eta$ can be chosen such that the left hand side is upper-bounded by $\epsilon>0$, which completes the proof.

We now substitute the construction \eqref{eq:etaLimitClaim2} into \eqref{eq:2ndOrderAmpl_farField} and perform the integrals. This results in the linear expansion
\begin{equation}\label{eq:FarFieldBogoTransf}
    \alpha_{\pmu}^{\rm out} \equiv \lim_{t\to\infty} \alpha_\pmu (t) e^{icpt} = \sum_\keps \left[ u_{\pmu,\keps}^{\rm out}\alpha^{\rm in}_\keps + v_{\pmu,\keps}^{\rm out}\alpha^{\rm in\dag}_\keps \right]
\end{equation}
with the coefficients
\begin{align}\label{eq:upk_App_farField}
  u_{\pmu,\keps}^{\rm out} = \frac{\sqrt{kp}}{L^3}
  (\ve_x \cdot \ve_{\keps}) (\ve_x \cdot \ve_{\pmu}) \dens_{\vk} \dens_{\vp} e^{i(\vk-\vp)\cdot\vr_0}
  \bigg[\underbrace{ \frac{1}{p-k-i0^+} - \frac{1}{p-k+i0^+} }_{= 2\pi i\delta(p-k)}\bigg] \chi(ck+i0^+) + \delta_{\pmu,\keps}, \qquad v_{\pmu,\keps}^{\rm out} = 0.
\end{align}
The response function $\chi(\omega)$ is defined in \eqref{eq:chi}. The energy condition $k=p$ ensures that all scattered field components remain far red-detuned with respect to the scatterer and the high-frequency cutoff, $cp\ll \omega_0$ and $pa_0 \ll 1$. Hence, we can set $\chi(ck+i0^+) = \chi(ck)$, neglect the $a_0$-terms in the denominator and carry out the continuum limit, which results in the scattering amplitudes \eqreff{FarFieldAmplitude} in the main text.

For an incident light pulse along $z$, $\alpha_\keps^{\rm in} = L^2 \alpha^{\rm in} (k) \delta_{\eps, 1} \delta_{k_x,0}\delta_{k_y,0}$, with polarization $\ve_\keps = \ve_x$ and amplitude density per unit area $\alpha^{\rm in} (k) = \alpha_{k\ve_z}^{\rm in}/L^2$, the total number of incident photons per unit area is $\Phi = (1/L^2) \sum_\keps |\alpha_\keps^{\rm in}|^2 \to (L^3/2\pi) \int dk |\alpha^{\rm in} (k)|^2 $. With the above approximations, the scattering amplitudes are simply $\alpha_\pmu^{\rm out} - \alpha_\pmu^{\rm in} =  ip \chi (cp) (\ve_x\cdot\ve_\pmu) e^{i(p\ve_z-\vp)\cdot \vr_0} \alpha^{\rm in} (p)$,
from which we obtain the total number of scattered photons,
\begin{align}
    N^{\rm sc} &=  \sum_\pmu |\alpha_\pmu^{\rm out} - \alpha_\pmu^{\rm in}|^2 \to  \left(\frac{L}{2\pi}\right)^3 \int d^3p \left| p\chi (cp) \alpha^{\rm in} (p) e^{i(p\ve_z - \vp)\cdot \vr_0} \right|^2 \sum_\mu (\ve_x\cdot\ve_\pmu)^2 \nonumber \\
    &= \left(\frac{L}{2\pi}\right)^3 \int_0^\infty dp \, p^4 | \chi (cp) \alpha^{\rm in} (p)|^2 \int d \Omega_{\vp} \left[1-(\ve_x\cdot\ve_{\vp})^2\right] \approx \frac{2}{3\pi} (\kin)^4 |\chi_0|^2 \Phi \equiv \sigma_{\rm tot} \Phi.
\end{align} 
In the final step, we have used that $\chi (cp) \approx \chi_0$ and $p\approx \kin$ for sufficiently narrow-band off-resonant incident light, which results in the total scattering cross section $\sigma_{\rm tot} = 2(\kin)^4|\chi_0|^2/3\pi$. Notice that our definition of the response $\chi_0$ corresponds to a polarizability of $2\epsilon_0 \chi_0$ in SI-units, which yields the well-known dipole scattering cross section.

\clearpage 

\section{Field Expectation Values}\label{app:FieldExpectationValues}

Here we verify the agreement between the phenomenological dipole radiation fields \eqreff{scattered_field} in the main text and the expectation values of the physical fields resulting from our quantum scattering model for any distance $\rho>0$ from the scatterer. To this end, we will evaluate the expectation values of the transverse field variables in the multipolar PZW gauge from the exact time-evolved expressions for the coherent mode amplitudes, as stated in the main text and derived in Supplementary Section~\ref{sec:TimeEvol}. 
We will carry out the calculation for a regularized dipole assuming $\rho \gg a_0$ and perform the point dipole limit $a_0 \to 0$ in the end.

The quantum operators of the transverse vector potential $\hatvA_T (\vr)$ and its (gauge-dependent) conjugate $\hatvPi (\vr)$ at the position $\vr = \vr_0 + \vrho$ of a detector pixel are expanded in terms of the plane-wave mode operators $\oa_\pmu$ in \eqref{eq:fieldquantization}. The time-dependent expectation values of the latter are the coherent amplitudes $\alpha_\pmu (t)$ in \eqref{eq:alpha_App} which, after inserting the expansion coefficients \eqref{eq:upk_App} and \eqref{eq:vpk_App}, can be split into an incident amplitude $\alpha_\pmu^{\rm in} e^{-icpt}$ and a scattered amplitude, $\alpha_\pmu^{\rm sc} (t) = \alpha_\pmu (t) - \alpha_\pmu^{\rm in} e^{-icpt}$. 
For consistency with the phenomenological setting, we shall now assume $\alpha_\pmu^{\rm in} = \alpha^{\rm in}\delta_{\pmu, \keps}$, corresponding to stationary off-resonant illumination by a single mode of wave vector $\vk = \kin\ve_z$, $c\kin < \omega$, and polarization $\ve_\keps = \ve_x$. Hence, the scattered amplitude simplifies to 
\begin{equation}\label{scAmplPlaneWave}
 \alpha_\pmu^{\rm sc} (t) = \frac{\sqrt{\kin p}}{L^3} e^{-i\vp\cdot\vr_0} \chi_0 (\ve_\pmu \cdot \ve_x) \dens_{\vp} \dens_{\vkin} \left[ \frac{\alpha^{\rm in}}{p-\kin -i0^+} e^{i\kin(z_0-ct)} - \frac{\alpha^{\rm in*}}{p+\kin} e^{-i\kin(z_0-ct)} \right],
\end{equation}
with the real-valued off-resonant polarizability $\chi_0 \equiv \chi (c\kin)$. 
Accordingly, the mean transverse vector potential $\la \hatvA_T (\vr) \ra_t$ splits into the incident $\vA_T^{\rm in} (\vr,t) =A^{\rm in} \ve_x e^{i\kin(z - ct)} + c.c.$, with $A^{\rm in} = \alpha^{\rm in}\sqrt{\hbar/2\epsilon_0 c \kin L^3}$, and the scattering component $\vA_T^{\rm sc} (\vr,t)$. In order to obtain the physical fields, we focus our attention on the conjugate, $\la \hatvPi (\vr)\ra_t = \partial_t \vA_T^{\rm in} (\vr,t) + \boldsymbol\Pi^{\rm sc} (\vr,t)$, which in the PZW gauge and away from the scatterer represents the negative electric field. The scattering contribution is 
\begin{align}
  \boldsymbol\Pi^{\rm sc} (\vr,t) &= -i\sqrt{\frac{\hbar c}{2\epsilon_0 L^3}}\sum_\pmu \sqrt{p} \ve_\pmu e^{i\vp \cdot \vr } \alpha_\pmu^{\rm sc} (t) + c.c. \nonumber \\
  &\xrightarrow{L\to\infty} -i c \kin \chi_0 \dens_{\vkin} \int \frac{d^3 p}{(2\pi)^3} p e^{i\vp\cdot\vrho} \dens_{\vp} \left[ \frac{A^{\rm in}e^{i\kin(z_0-ct)}}{p-\kin-i0^+} - \frac{A^{\rm in*}e^{-i\kin(z_0-ct)}}{p+\kin} \right] \sum_\mu \ve_\pmu (\ve_\pmu \cdot\ve_x) + c.c. \nonumber \\
  &= -i c \kin \chi_0 \dens_{\vkin} \int_0^\infty \frac{d p}{2\pi^2} p^3 \dens_{\vp} \left[ \frac{A^{\rm in}e^{i\kin(z_0-ct)}}{p-\kin-i0^+} - \frac{A^{\rm in*}e^{-i\kin(z_0-ct)}}{p+\kin} \right] \underbrace{\int \frac{d\Omega}{4\pi} e^{i\vp\cdot\vrho} \sum_\mu \ve_\pmu (\ve_\pmu \cdot\ve_x)}_{\equiv \vF (p\rho)} + c.c. \nonumber \\
  &= -i c \kin \chi_0 \dens_{\vkin} \int_{-\infty}^\infty \frac{d p}{2\pi^2} p^3 \vF(p\vrho) \dens_{\vp}  \left[ \frac{A^{\rm in}e^{i\kin(z_0-ct)}}{p-\kin-i0^+} - \frac{A^{\rm in*}e^{-i\kin(z_0-ct)}}{p+\kin-i0^+} \right] \nonumber \\
  &= -i c \kin \chi_0 A^{\rm in} \dens_{\vkin}  e^{i\kin(z_0-ct)} \int_{-\infty}^\infty \frac{d p}{2\pi^2} \frac{p^3 \vF(p\vrho) \dens_{\vp}}{p-\kin-i0^+} + c.c. = -\vE^{\rm sc} (\vr,t) . \label{eq:Pisc}
\end{align}
From the third to the fourth line, the complex conjugate is absorbed by extending the $p$-integral to $-\infty$. Recalling that the $\ve_\pmu$ are two basis vectors orthogonal to $\ve_{\vp} = \vp/p$, we have $\sum_\mu \ve_\pmu (\ve_x\cdot \ve_\pmu) = \ve_x - \ve_{\vp} (\ve_{\vp}\cdot\ve_x)$. Let us now define the solid angle with respect to the polar axis $\ve_{\vrho} = \vrho/\rho$ and the two azimuthal axes $\ve_1,\ve_2$, such that $\ve_{\vp} = \cos \vartheta \ve_{\vrho} + \sin\vartheta (\cos\varphi \ve_1 + \sin\varphi \ve_2)$ and $\ve_x = \cos\gamma \ve_{\vrho} + \sin\gamma \ve_1$. The angular part of the integral then simplifies as
\begin{align}
    \vF(p\vrho) &= \int \frac{d\Omega}{4\pi} e^{\vp\cdot\vrho} \left[ \ve_x - \ve_{\vp} (\ve_{\vp}\cdot\ve_x)\right] = \int \frac{d\cos\vartheta d \varphi}{4\pi} e^{p\rho \cos\vartheta } \left[ \cos \gamma (1-\cos^2\vartheta)\ve_{\vrho} + \sin\gamma (1-\sin^2\vartheta \cos^2\varphi)\ve_1  \right] \nonumber \\
    &= (\ve_{\vrho} \times \ve_x) \times \ve_{\vrho} \frac{\sin p\rho}{p\rho} + \left[ \ve_x - 3\ve_{\vrho} (\ve_x \cdot \ve_{\vrho} ) \right] \frac{p\rho \cos p\rho - \sin p\rho}{(p\rho)^3} ,
    \label{VectorPotentialAngularIntegral}
\end{align}
which assumes a finite value also at $p=0$.
Analogously, the magnetic field is
\begin{align}
    \vB^{\rm sc}(\vr, t) &= \nabla \times \vA_T(\vr, t) = i\sqrt{\frac{\hbar}{2\epsilon_0 c L^3}} \sum_\pmu \sqrt{p} \ve_{\vp} \times \ve_{\pmu} e^{i\vp\cdot\vr} \alpha_\pmu^{\rm sc} (t) + c.c.
    \nonumber \\ &\xrightarrow[L\rightarrow\infty]{} i \kin \chi_0 A^{\rm in} \dens_{\vkin} e^{i\kin(z_0-ct)} \int_{-\infty}^\infty \frac{d p}{2\pi^2} \frac{p^2 \nabla \times \vF(p\vrho) \dens_{\vp}}{p-\kin-i0^+} + c.c.
    \label{eq:Bsc}
\end{align}
The curl of $\vF$ is \begin{equation}
    \nabla \times \vF(p\vrho) = \int \frac{d\Omega}{4\pi} e^{\vp\cdot\vrho} \left[ \vp \times \ve_x \right] = p\int \frac{d\cos\vartheta d \varphi}{4\pi} e^{p\rho \cos\vartheta } \sin\gamma \cos\vartheta\, \ve_2
     = p\, \ve_{\vrho} \times \ve_x \frac{\sin p\rho - p\rho\cos p\rho}{(p\rho)^2}.
    \label{MagneticFieldPotentialAngularIntegral}
\end{equation}
We can now carry out the remaining $p$-integrals in \eqref{eq:Pisc} and \eqref{eq:Bsc} with the help of the residue theorem. To this end, we must express $\sin p\rho = (e^{ip\rho}-e^{-ip\rho})/2i$ and $\cos p\rho = (e^{ip\rho}+e^{-ip\rho})/2$ in $\vF$. Since $\rho>0$, the integration contour must be closed in the complex upper half-plane for the $e^{ip\rho}$ terms and in the lower half-plane for the $e^{-ip\rho}$ terms. Writing $i0^+ = i\eta$ in terms of an infinitesimal $\eta>0$, the integrand has a pole at $p=\kin+i\eta$ in the upper half-plane, while the regularisation factor contributes two additional poles at $p = \pm i/a_0$.
We arrive at
\begin{align}
    \int_{-\infty}^\infty \frac{d p}{2\pi^2} \frac{p^3 \vF(p\vrho) \dens_{\vp}}{p-\kin-i0^+} &= \frac{(\kin)^3 \dens_{\vk}}{2\pi} \left\{
    \left[ \left(\frac{4\rho}{a_0(a_0\kin)^2} + \frac{\rho-a_0}{a_0} \right)e^{-2\rho/a_0} + e^{i\kin\rho} \right] 
    \frac{(\ve_{\vrho} \times \ve_x) \times \ve_{\vrho}}{\kin\rho} 
    \right. \nonumber \\
    & \quad + \left[ i\kin\rho
    \left( i \left(\frac{ \kin(a_0-2\rho)}{4} - \frac{a_0 + 2\rho}{a_0^2\kin} \right)e^{-2\rho/a_0} + e^{i\kin\rho} \right) \right. \nonumber \\
    & \quad + \left. \left. \left( \left(\frac{\rho}{a_0} + 1 + \frac{(a_0 \kin)^2\rho}{4a_0}\right) e^{-2\rho/a_0} - e^{i\kin\rho} \right)
    \right] \frac{ \ve_x - 3\ve_{\vrho} (\ve_x \cdot \ve_{\vrho} ) }{(\kin\rho)^3} \right\}, 
    \\
    \int_{-\infty}^\infty \frac{d p}{2\pi^2} \frac{p^2 \nabla \times \vF(p\vrho) \dens_{\vp}}{p-\kin-i0^+} 
    &= \frac{(\kin)^3 \dens_{\vk}}{2\pi}\left[ 
    -\left( i \left(\frac{ \kin(a_0-2\rho)}{4} - \frac{a_0 + 2\rho}{a_0^2\kin} \right)e^{-2\rho/a_0} + e^{i\kin\rho} \right)
    \right.\nonumber \\ & \left. \quad 
    - i\kin\rho
     \left( \left(\frac{\rho}{a_0} + 1 + \frac{(a_0 \kin)^2\rho}{4a_0}\right) e^{-2\rho/a_0} - e^{i\kin\rho} \right)
    \right] \frac{\ve_{\vrho} \times \ve_x}{(\kin\rho)^2}
\end{align}
Inserting this into \eqref{eq:Pisc} and \eqref{eq:Bsc} yields explicit expressions for the regularized scattering fields at distances $\rho \gg a_0$ away from the scatterer: 
\begin{align}\label{eq:Esc_regularized}
    \vE^{\rm sc} (\vr,t) = -\vPi^{\rm sc}(\vr, t) &= \frac{i c (\kin)^4 \chi_0 A^{\rm in} \dens_{\vkin} }{2\pi} e^{i\kin(\rho+z_0-ct)} \left\{
    \left[ \left(\frac{4\rho}{a_0(a_0\kin)^2} + \frac{\rho-a_0}{a_0} \right)e^{-2\rho/a_0} + e^{i\kin\rho} \right] 
    \frac{(\ve_{\vrho} \times \ve_x) \times  \ve_{\vrho}}{\kin\rho} \right. \nonumber \\ 
    &\quad + \left[ i\kin\rho
    \left( i \left(\frac{ \kin(a_0-2\rho)}{4} - \frac{a_0 + 2\rho}{a_0^2\kin} \right)e^{-2\rho/a_0} + e^{i\kin\rho} \right) \right. \nonumber \\
    &\quad + \left. \left. 
    \left( \left(\frac{\rho}{a_0} + 1 + \frac{(a_0 \kin)^2\rho}{4a_0}\right) e^{-2\rho/a_0} - e^{i\kin\rho} \right)
    \right] \frac{ \ve_x - 3\ve_{\vrho} (\ve_x \cdot \ve_{\vrho} ) }{(\kin\rho)^3} \right\}, \\ 
    \vB^{\rm sc} (\vr,t) &= \frac{i (\kin)^4 \chi_0 A^{\rm in} \dens_{\vkin} }{ 2\pi} e^{i\kin(z_0-ct)} \left\{ -\left( i \left(\frac{ \kin(a_0-2\rho)}{4} - \frac{a_0 + 2\rho}{a_0^2\kin} \right)e^{-2\rho/a_0} + e^{i\kin\rho} \right)
    \right.\nonumber \\ & \left. \quad 
    - i\kin\rho
     \left( \left(\frac{\rho}{a_0} + 1 + \frac{(a_0 \kin)^2\rho}{4a_0}\right) e^{-2\rho/a_0} - e^{i\kin\rho} \right) \right\} \frac{\ve_{\vrho} \times \ve_x}{(\kin\rho)^2} .
    \label{eq:Bsc_regularized}
\end{align}
If we set $A^{\rm in} \equiv -i E^{\rm in}/c\kin$ and go back to the ideal point dipole case, $a_0 \to 0$, we retrieve the phenomenological expressions \eqreff{scattered_field} from the main text.

\subparagraph{Finite dipole polarization density}

\new{%
By imposing a UV regularization of the dipole scattering field in terms of the length scale parameter $a_0>0$, we have ensured that the near-field QFI would not diverge. Here we provide the intuitive physical meaning of $a_0$: it represents the radial extension of a dipole polarization density $\dens (\vr) = e^{-2r/a_0}/(\pi a_0^3)$ that describes the dipole scatterer in the regime $a_0 \kin \ll 1 $. 
Explicitly, we claim that the canonical scattering field \eqref{eq:Pisc} is, to a good approximation, given by the integrated scattering field emanating from the density $\dens(\vr)$,
\begin{equation}\label{PiConvolution}
    \vPi^{\rm sc} (\vr,t) \stackrel{a_0\kin \ll 1}{\approx} \vPi_\dens^{\rm sc} (\vr,t) := \int d^3r' e^{i \vkin \cdot \vr'} \dens(\vr') \tildevPi^{\rm sc} (\vr - \vr',t) + c.c. 
\end{equation}
Here, the factor $e^{i\vkin\cdot \vr'}$ accounts for the path lengths traveled by the incident light wave to the locations $\vr'$ occupied by the polarization density, and the complex point-dipole scattering field reads as
\begin{equation}\label{PointDipolePi}
    \tildevPi^{\rm sc} (\vr,t) = -i\sqrt{\frac{\hbar c}{2\epsilon_0 L^3}}\sum_\pmu \sqrt{p} \ve_\pmu e^{i\vp \cdot \vr } \tilde\alpha_\pmu^{\rm sc} (t), \qquad  \tilde\alpha_\pmu^{\rm sc} (t) = \lim_{a_0 \to 0} \alpha_\pmu^{\rm sc} (t).
\end{equation}
Inserting this plane-wave expansion into \eqref{PiConvolution}, we can express the integrated scattering field in terms of the Fourier components of the polarization density, $ \dens_{\vk} = \int d^3r \, e^{-i\vk\cdot\vr} \dens (\vr) = 1/[1+(a_0k/4)^2]^2$; namely, 
\begin{align}
    \vPi_\dens^{\rm sc} (\vr,t) &= -i \sqrt{\frac{\hbar c}{2\epsilon_0 L^3}} \sum_{\pmu} \sqrt{p} \ve_\pmu e^{i\vp \cdot \vr } \dens_{\vp-\vkin} \tilde\alpha_\pmu^{\rm sc} (t) + c.c.
\end{align} 
Now we can simply notice that, to lowest order, $  \dens_{\vp-\vkin}\tilde\alpha_\pmu^{\rm sc} (t) =  \dens_{\vp}\tilde\alpha_\pmu^{\rm sc} (t) +\mathcal O (a_0\kin) = \alpha_\pmu^{\rm sc} (t) + \mathcal O (a_0\kin)$, which proves our claim in \eqref{PiConvolution}. 

We remark that Ref.~\cite{Stokes2022ImplicationsElectrodynamics} employs a simpler UV regularization in terms of the Fourier components $ \dens_{\vk}' = 1/[1+(a_0k)^2]$, which amounts to a polarization density $ \dens'(\vr) = e^{-r/a_0}/4\pi a_0^2 r$. Unfortunately, this density diverges at the origin and thus would retain a logarithmic divergence of the near-field QFI, which is why we use the bounded $\dens (\vr)$ instead.
}

\clearpage 
\twocolumngrid

\section{Gauge Relativity of the quantum Fisher information}
\label{sec:GaugeInvariance}

In the main text, we evaluated the information content of the quantum state of the transverse light field at a given time $t$ about the dipole scatterer polarizability and position, $\vtheta = (\chi_0,\vr_0)$, as measured by the quantum Fisher information (QFI) matrix $\mathcal J (\vtheta,t)$. Here we argue that this QFI is in general not invariant under the choice of electromagnetic gauge. However, for a standard dipole detector model, there is a preferred gauge---the multipolar PZW gauge we assume in the main text---in which the state of the transverse field degrees of freedom captures all the detectable information the scatterer transmits to the field. The QFI in this gauge is thus optimal compared to that in other gauges, assuming the same detector model.

\subparagraph{Unitary invariance of the QFI} 

In order to understand the gauge relativity of the QFI matrix, recall its formal definition given in Supplementary Section \ref{sec:ParameterEstimationIntro}. In our setting, the initial state of the field and scatterer $\varrho(t_0)$ acquires information about the unknown scatterer parameters $\vtheta$ through its time evolution under the light-matter Hamiltonian $\oH = \oH(\vtheta)$: $\varrho(t,\vtheta) = e^{-i\oH(\vtheta)(t-t_0)/\hbar} \varrho (t_0) e^{i\oH(\vtheta)(t-t_0)/\hbar} $. The QFI matrix $\mathcal J$ of this state is then defined as the optimum of FI matrices $\mathcal I$ taken over all possible POVMs on the state. 
It follows immediately that the QFI matrix is unchanged under unitary transformations, $\varrho (t,\vtheta) \to \oU \varrho (t,\vtheta) \oU^\da$, which could be used to change the frame or representation of the quantum system. Crucially, this assumes that the unitaries themselves do not depend on the parameters $\vtheta$ to be estimated.

\subparagraph{Gauge transformations}

Here, the quantum system is a dipole scatterer (modeled as a harmonic oscillator) interacting with the electromagnetic radiation field, and the initial state at $t_0 \to -\infty$ describes an incident coherent pulse of probe light and the scatterer in its ground state. However, the exact representation of the quantum state $\varrho_g$ and the Hamiltonian $\oH_g(\vtheta)$ of scatterer and field depends on the chosen electromagnetic gauge $g$~\cite{Stokes2022ImplicationsElectrodynamics}. One typically starts with the minimal coupling Hamiltonian in the Coulomb gauge $g'$ and then switches to a more convenient gauge $g$ by means of a unitary gauge fixing transformation $\oU_{gg'}$. The quantum state transforms as $\varrho_{g'} \to \varrho_g = \oU_{gg'}\varrho_{g'}\oU_{gg'}^\da$. Gauge relativity of the QFI in our setting can be attributed to two problems.

Firstly, the most expedient gauges in the case of a dipole scatterer depend on its position $\vr_0$. In particular, the multipolar PZW gauge, for which the scatterer-field interaction reduces to the (regularized) textbook form \eqref{eq:HI_app_regularized} of a dipole Hamiltonian, is fixed by 
\begin{equation}\label{eq:U_CoulombToPZW}
    \oU_{gg'} = \exp \left[ - \frac{i}{\hbar} \frac{1}{L^3}\sum_{\vk} \frac{q \hatvr_e\cdot \hatvA_{T\vk}}{[1+\tfrac{1}{4}(a_0k)^2]^2} e^{-i\vk\cdot \vr_0} \right] .
\end{equation}
Since it is explicitly determined by the parameters $\vr_0$ that we seek to estimate, we cannot expect the same QFI for $\varrho_g$ and $\varrho_{g'}$. 

Secondly, we have no direct access to the state of the scatterer here, but only to the field through photodetection. Hence, the relevant QFI in our setting is that of the reduced state of the (transverse) field degrees of freedom. Given that gauge fixing transformations, whether they depend on $\vr_0$ or not, may correlate and exchange information between the scatterer and the field, the QFI of the reduced field state may change, too. 

\subparagraph{PZW versus Coulomb gauge}

As an illustration of the gauge relativity, we compare the PZW gauge employed in this work with the Coulomb gauge. Once a gauge $g$ is fixed, the transverse field excitations are quantized by expanding the (gauge-invariant) transverse vector potential $\hatvA_{T} (\vx)$ and its (gauge-variant) canonical conjugate $\hatvPi_g(\vx)$ in a chosen mode basis and taking the expansion coefficients as the 'position' and 'momentum' quadratures. Here, we quantize the free-space field in the basis of plane waves, according to the rule \eqref{eq:fieldquantization} in the PZW gauge, and the resulting photon degrees of freedom are represented by the ladder operators 
\begin{align}\label{LadderOpFromFields}
    \oa_{\keps,g} &= \sqrt{\frac{\epsilon_0}{2\hbar}} \left(
    \sqrt{ck} \hatvA_{T\vk} + \frac{i}{\sqrt{ ck}} \boldsymbol{\hat{\Pi}}_{\vk,g}
    \right) \cdot \ve_\keps, \nonumber \\
    \hatvA_{T\vk} &= \int_{L^3} \frac{d^3 x}{\sqrt{L^3}} \hatvA_{T} (\vx) e^{-i\vk\cdot\vx},  \\
    \hatvPi_{\vk,g} &= \int_{L^3} \frac{d^3 x}{\sqrt{L^3}} \hatvPi_{g} (\vx) e^{-i\vk\cdot\vx}. \nonumber 
\end{align}
However, the meaning of a photon differs from gauge to gauge. The photon annihilation operator $\oa_\keps \equiv \oa_{\keps,g}$ we are using in the PZW gauge transforms under \eqref{eq:U_CoulombToPZW} back into the Coulomb gauge as 
\begin{align}
    \oa_{\keps}' &= \oU_{gg'}^\da \oa_{\keps} \oU_{gg'} = \oa_{\keps} - \frac{i q \ovr_e \cdot \ve_\keps }{\sqrt{2\eps_0 \hbar c k L^3}} \frac{e^{-i\vk\cdot\vr_0}}{[1+\tfrac{1}{4}(a_0k)^2]^2}  \nonumber \\
    &= \oa_{\keps} - \frac{i d_0 \ve_x \cdot \ve_\keps }{\sqrt{2\eps_0 \hbar c k L^3}} \frac{e^{-i\vk\cdot\vr_0}}{[1+\tfrac{1}{4}(a_0k)^2]^2} (\ob + \ob^\da).
    \label{eq:akeps_trafo_PZWtoCoulomb}
\end{align}
This operator represents the same expectation value, $\tr [\oa_{\keps}' \varrho_{g'}] = \tr [\oa_{\keps} \varrho_{g}]$, but it now acts on both the field and the scatterer subsystem. On the other hand, the average coherent amplitude for the respective Coulomb-gauge photon mode would be 
\begin{equation}
    \alpha_\keps' \equiv \tr [ \oa_{\keps} \varrho_{g'}] = \alpha_\keps - \frac{i d_0 \ve_x \cdot \ve_\keps }{\sqrt{2\eps_0 \hbar c k L^3}} \frac{e^{-i\vk\cdot\vr_0} (\beta + \beta^*)}{[1+\tfrac{1}{4}(a_0k)^2]^2} ,
\end{equation}
where $\alpha_\keps$ and $\beta$ are the coherent amplitudes of the field mode and the scatterer in PZW gauge. The Coulomb-gauge amplitudes are also linear combinations of the incident $\alpha_\keps^{\prime \rm in } = \alpha_\keps^{\rm in}$. By expanding the operators in \eqref{eq:akeps_trafo_PZWtoCoulomb} according to \eqref{eq:alpha_App}, and with help of identity \eqref{eq:beta_relation}, we find that their expansion coefficients can be given in terms of the PZW-gauge coefficients simply by $u_{\pmu,\keps}' = -(k/p)u_{\pmu,\keps}$ and $v_{\pmu,\keps}' = -(k/p)v_{\pmu,\keps}$.

Similarly, we can use the transformation rule \eqref{eq:akeps_trafo_PZWtoCoulomb} to calculate the covariance matrix blocks $\Xi_{g'},\Upsilon_{g'}$ of the field degrees of freedom in the Coulomb gauge, as well as the derivatives with respect to the parameters $\vtheta$ as in Supplementary Section~\ref{sec:QFI_Appendix}. This allows us to re-evaluate for our scattering problem the QFI in the reduced field state, as seen from the Coulomb gauge. 
We do not repeat the full calculation, since it proceeds along the same lines as \eqref{eq:dAlphaDR_eq1}-\eqref{eq:f123}. We simply state the relevant frequency integrals $f_{1,2,3}'$, which differ from the $f_{1,2,3}$ in \eqref{eq:f123} by a factor $k/p$ inside the $k$-integral, and by 
an overall sign:

\begin{widetext}
\begin{equation}\begin{aligned}\label{eq:f123Coulomb}
f_1'(p,t) &= \frac{p^{-1/2}}{[1+\tfrac{1}{4}(a_0p)^2]^2}\int_0^\infty \frac{dk}{2\pi} \frac{k^{5/2}}{[1+\tfrac{1}{4}(a_0k)^2]^2} \left[
-\frac{\chi( ck-i0^+) \alpha^{\rm in *}(k,t)}{k+p} +
\frac{\chi( ck+i0^+)\alpha^{\rm in}(k,t)}{k-p+i0^+}
\right],
\\\\
f_2'(p,t) &=
\frac{p^{1/2}}{[1+\tfrac{1}{4}(a_0p)^2]^2} \int_0^\infty \frac{dk}{2\pi} \frac{k^{3/2}}{[1+\tfrac{1}{4}(a_0k)^2]^2} \left[
-\frac{\chi( ck-i0^+) \alpha^{\rm in *}(k,t)}{k+p} -
\frac{\chi( ck+i0^+) \alpha^{\rm in}(k,t)}{k-p+i0^+}
\right],
\\\\
f_3'(p,t) &= \frac{p^{-1/2}}{[1+\tfrac{1}{4}(a_0p)^2]^2} \int_0^\infty \frac{dk}{2\pi} \frac{k^{3/2}}{[1+\tfrac{1}{4}(a_0k)^2]^2} \left[
-\frac{\partial\chi(ck-i0^+)}{\partial \chi_0} \frac{\alpha^{\rm in *}(k,t)}{k+p} -
\frac{\partial\chi(ck+i0^+)}{\partial \chi_0} \frac{\alpha^{\rm in}(k,t)}{k-p+i0^+}
\right].
\end{aligned}\end{equation}
\end{widetext}

\begin{figure}[b]
  \begin{overpic}[width=\linewidth]{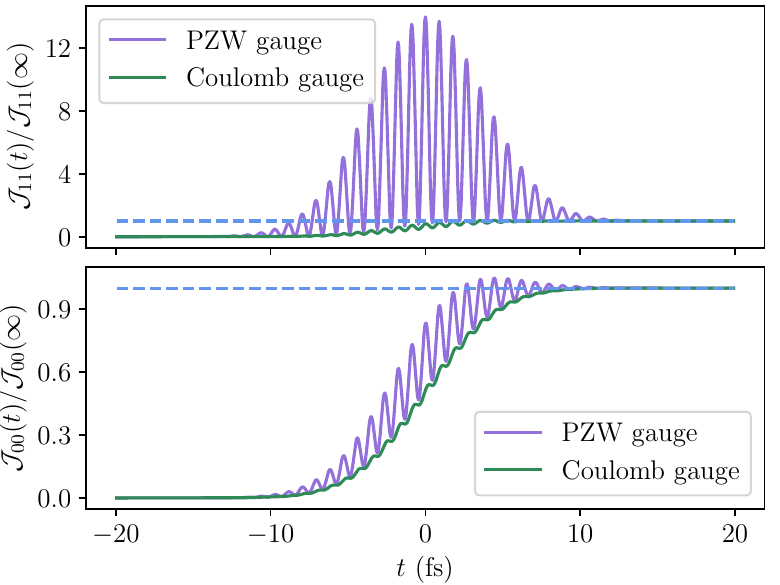}
    \put (85, 30) {\large (b)}
    \put (85, 65) {\large (a)}
  \end{overpic}
  \caption{QFI for estimating $\theta_1 = x_0$ and $\theta_0 = \chi_0$, for a scatterer with radius $a_0 = \lambdain/30$.
  }
  \label{fig:QFI_GaugeCompare}  
\end{figure}

\begin{figure}[t]
  \begin{overpic}[width=\linewidth]{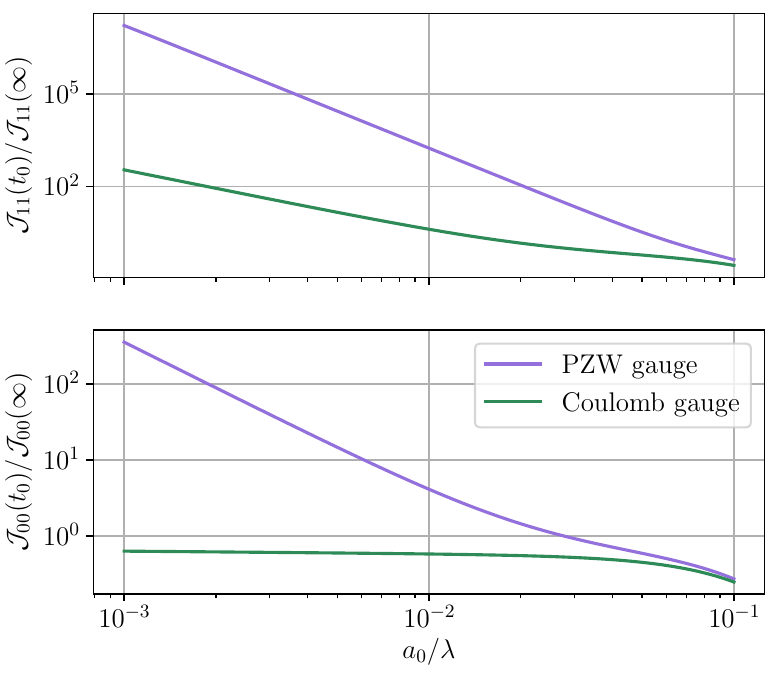}
    \put (17, 30) {\large (b)}
    \put (17, 75) {\large (a)} 
  \end{overpic}
  \caption{Near-field QFI as a function of $a_0$ for a fixed polarizability $\chi_0 = \SI{13.0}{nm^3}$ at $\lambdain=\SI{532}{nm}$. 
  The near-field oscillations peak at $t_0=0$.
  Upper panel is for estimating $x_0$, lower panel for $\chi_0$. The $(\lambdain/a_0)^4$ and $(\lambdain/a_0)^2$ powers in the PZW gauge reduce to $(\lambdain/a_0)^2$ and $(\lambdain/a_0)^0$ in the Coulomb gauge.
  }
  \label{fig:PowerLaws_GaugeCompare}
\end{figure}

Figure \ref{fig:QFI_GaugeCompare} plots an exemplary comparison of the QFI matrix elements in the PZW (purple line) and the Coulomb gauge (green) as a function of time, associated to the parameters (a) $\theta_1 = x_0$ and (b) $\theta_0 = \chi_0$. The purple lines match those of Fig.~\reff{fig:QFI} in the main text, which uses the same settings. In both gauges, the QFI oscillates twice per optical cycle.
While the overall buildup over time can be observed in both gauges, with the same asymptotic far-field value, the transient near-field values in the PZW gauge clearly exceed those of the Coulomb gauge. In Figure \ref{fig:PowerLaws_GaugeCompare}, we plot the corresponding peak values of the QFI when the probe pulse hits the scatterer at $t=0$, as a function of the scatterer size $a_0$. In the PZW gauge, the peak QFI for (a) $x_0$ and (b) $\chi_0$ diverges like the fourth and the second power of $\lambdain/a_0$, respectively. In the Coulomb gauge, on the other hand, we find a divergence with only the second power in (a), and a saturation in (b). Clearly, the transverse field degrees of freedom in the PZW gauge learn more about the scatterer than in the Coulomb gauge. In the following, we will argue why the quantum state of the transverse field in the PZW gauge carries the most information about the scatterer to a photodetector, as compared to the Coulomb or other intermediate gauges~\cite{Stokes2022ImplicationsElectrodynamics}.

\subparagraph{Scatterer-detector interaction in the PZW gauge}

Ultimately, physical observables must always be gauge-invariant, and so must be parameter estimation based on them. This calls for a physical model for photodetection, determining which exact POVM measurement it represents in a given gauge. (Working out a realistic detector model is beyond the scope of this work.) At the same time, the purpose of the QFI is to provide an upper bound on how much information about certain parameters can at best be extracted from a quantum state by any measurement. We now show that the PZW gauge should give the greatest upper bound on the information that can possibly be extracted from photodetectors---when they are also modeled in the usual manner as (regularized) dipoles. 

To this end, consider multiple dipoles described by bound quantum charges $q_c\equiv q$ with canonical coordinates $(\ovr_c,\ovp_c)$, oscillating around the respective equilibrium positions $\vr_{0,c}$. We shall label by $c=0$ our scatterer of interest, $(\ovr_0,\ovp_0) \equiv (\ovr_e,\ovp_e)$ and $\vr_{0,0} \equiv \vr_0$, while $c>0$ represent the detector dipoles. The full Hamiltonian of all these dipoles and the field in the PZW-gauge is then the multiparticle equivalent of \eqref{eq:TotalH_Multipolar}:
\begin{align}\label{eq:TotalH_ManyParticle}
  \hat H_{\rm tot} &= \sum_c \left\{ \frac{[\hatvp_c - q\hatvA(\hatvr_c+\vr_{0,c})]^2}{2m} + U_c(\hatvr_c) \right\} +  V_{\rm self} \nonumber \\
  & + \frac{q^2}{4\pi \eps_0} \sum_{c<d} \left[
  \frac{\hatvr_c\cdot\hatvr_d}{\rho_{cd}^3} - \frac{3(\hatvr_c\cdot\vrho_{cd})(\hatvr_d\cdot\vrho_{cd})}{\rho_{cd}^5} \right]  \\
  & + \frac{\eps_0}{2}\int d^3 x \left[ \left(\hatvPi + \frac{1}{\eps_0} \hatvP_T \right)^2 + c^2(\nabla \times \hatvA_T)^2 \right]. \nonumber 
\end{align}
Here, $V_{\rm self}$ subsumes all (infinite) self-interaction terms. The $U_c$ in the first line represent the individual trapping potentials of each bound charge while the second line corresponds to the static dipole-dipole interaction between pairs of them, denoting the distance as $\vrho_{cd} = \vr_{0,c}-\vr_{0,d}$. The third line is the contribution of the transverse field, which includes the dipole-field coupling through the transverse polarization. In the regularized dipole approximation analogous to \eqref{eq:TransvP_Regularized}, using the same length scale $a_0$ for all dipoles, the Fourier components of the transverse polarization read as
\begin{equation}\label{TransvP_ManyElectron}
  \hatvP_{T\vk} = \frac{-q}{[1+\tfrac{1}{4}(a_0k)^2]^2}\sum_c e^{-i\vk\cdot\vr_{0,c}}\sum_\eps \ve_\keps (\ve_\keps\cdot\hatvr_c).
\end{equation} 
Consistently, we can also approximate $\hatvA(\hatvr_c+\vr_{0,c}) \approx 0$, reducing the Hamiltonian to 
\begin{equation}
  \hat H_{\rm tot} = \sum_c \hat H_c + V_{\rm self} + \hat H_I + \frac{\eps_0}{2} \int d^3 x [\hatvPi^2 + c^2 (\nabla \times \hatvA_T)^2],
\end{equation} 
where $\oH_c = \ovp_c^2/2m + U_c(\ovr_c)$. All dipole-dipole and dipole-field interaction 
terms are subsumed under 
\begin{align}\label{ManyParticleInteraction}
  \hat H_I &=  \frac{q^2}{4\pi \eps_0} \sum_{c<d} \left[
  \frac{\hatvr_c\cdot\hatvr_d}{\rho_{cd}^3} - \frac{3(\hatvr_c\cdot\vrho_{cd})(\hatvr_d\cdot\vrho_{cd})}{\rho_{cd}^5} \right] \nonumber \\
  &\quad + \frac{1}{2}\int d^3 x \left[ 2\hatvP_T \cdot \hatvPi + \frac{1}{\eps_0} \hatvP_T^2 \right].
\end{align}
In the second line, we have the coupling between the dipoles and the transverse field quadrature $\hatvPi$ as well as another dipole-dipole coupling term. Using Parseval's identity to express the latter term in Fourier space in the continuum limit,
\begin{align}\label{eq:PolarizIntegralParseval}
    \frac{1}{\eps_0}\int d^3 x \, \hatvP_T^2(\vx) &= \frac{1}{\eps_0 L^3} \sum_{\vk} \sum_{i=1}^3 |\ve_i\cdot\hatvP_{T\vk}|^2 \\
    &\xrightarrow[L\to\infty]{} \frac{1}{\eps_0} \int \frac{d^3k}{(2\pi)^3} \sum_{i=1}^3 |\ve_i\cdot\hatvP_{T\vk}|^2 \nonumber
\end{align}
we can plug in \eqref{TransvP_ManyElectron} to get 
\begin{align} \label{eq:Polariz_kSpace}
    &\sum_{i=1}^3 |\ve_i\cdot\hatvP_{T\vk}|^2
    = q^2\sum_{i=1}^3 \sum_{c,d} e^{i\vk \cdot \vrho_{cd}} \nonumber \\
    &\quad\times \sum_{\eps,\mu=1,2} \frac{(\ve_i\cdot\ve_\keps)(\ve_i\cdot\ve_{\vk\mu})(\hatvr_c\cdot\ve_\keps)(\hatvr_d\cdot\ve_{\vk\mu})}{[1+\tfrac{1}{4}(a_0k)^2]^4} \\
    &= q^2 \sum_{c,d} e^{i\vk\cdot \vrho_{cd}} \sum_{\eps=1,2} \frac{(\hatvr_c\cdot\ve_\keps)(\hatvr_d\cdot\ve_\keps)}{[1+\tfrac{1}{4}(a_0k)^2]^4}. \nonumber
\end{align}
Recall that the $\ve_\keps$ are two field polarization unit vectors orthogonal to $\vk$. We now bring in the assumption that the extent $a_0$ of the dipoles is much smaller than the distances $\rho_{c\neq d}$ between them. Therefore, the exponential $e^{i\vk\cdot \vrho_{c\neq d}}$ oscillates very rapidly in the relevant $k$-values at which the regularizing term $k a_0$ becomes appreciable, and we can neglect the latter. Substituting back into \eqref{eq:PolarizIntegralParseval} and omitting the self-interaction terms $c=d$, we obtain
\begin{align}\label{eq:PolarizIntegralParseval3}
  &\frac{1}{2\eps_0}\int d^3 x \boldsymbol{\hat P}_T^2(\vx) \\
  =\, & \frac{q^2}{\eps_0} \sum_{c<d} \int \frac{d^3k}{(2\pi)^3} e^{i\vk\cdot \vrho_{cd}} \sum_{\eps=1,2} (\hatvr_c\cdot\ve_\keps)(\hatvr_d\cdot\ve_\keps) \nonumber \\
  =\, & -\frac{q^2}{4\pi \eps_0} \sum_{c<d} \left[
  \frac{\hatvr_c\cdot\hatvr_d}{\rho_{cd}^3} - \frac{3(\hatvr_c\cdot\vrho_{cd})(\hatvr_d\cdot\vrho_{cd})}{\rho_{cd}^5} \nonumber
  \right] .
\end{align}
This is exactly the negative of the dipole-dipole potential between the bound charges $c$ and $d$, and cancels with the first line in \eqref{ManyParticleInteraction}. We are left with an interaction solely mediated by the transverse field,
\begin{equation}
  \hat H_I = \int d^3x \, \hatvPi(\vx)\cdot\hatvP_T(\vx),
\end{equation}
which we can evaluate by expanding $\hatvPi$ in plane-wave modes and modeling the charges as harmonic oscillators as we did in Supplementary Section~\ref{sec:InteractionHamiltonian}. 

The result proves that there is (to a good approximation) no \new{separate} dipole-dipole interaction \new{term} between the scatterer and the detector in the PZW gauge---a distinguishing feature compared to the Coulomb or other intermediate gauges. All the information that the scatterer broadcasts into its surroundings is transmitted to the detector dipoles via the transverse field, and thus captured by the QFI of the reduced field state in this gauge. \new{In particular, the near-field dipole-dipole interaction between the scatterer and the detector is also mediated by the transverse field in the PZW gauge. In the Coulomb gauge, on the other hand, it is the longitudinal field that carries part of the near-field information, which shows up here as the separate dipole-dipole interaction term, thus depleting the QFI of the transverse field state.}

We remark that, if the assumption $\rho_{cd} \gg a_0$ does not hold, we cannot approximate the regularizing denominator in \eqref{eq:Polariz_kSpace} by unity, and  \eqref{eq:PolarizIntegralParseval3} is no longer valid. Hence, the QFI of the reduced field state only characterizes measurements made by detectors that do not overlap with the scatterer region of radius $a_0$. That is to say, detection schemes in such close vicinity are not subject to the quantum Cram\'{e}r-Rao bound evaluated here.

\end{document}